\documentclass[a4paper,8pt]{article}
\usepackage[english]{babel}
 
\usepackage[utf8x]{inputenc}
\usepackage[T1]{fontenc}
\usepackage{rotating}
\usepackage{comment}
\RequirePackage[colorlinks=false,citecolor=black,urlcolor=black,linkcolor=black,hidelinks]{hyperref}
\usepackage{dblfloatfix}
\usepackage[a4paper,top=2cm,bottom=2cm,left=2.5cm,right=2.5cm,marginparwidth=1.5cm]{geometry}
\usepackage{amsmath,amsthm,enumitem}
\usepackage{amsfonts}
\usepackage{subcaption}
\usepackage{graphicx}
\usepackage[superscript,biblabel,nomove]{cite} 
\usepackage{mathptmx}
\usepackage{tocloft}
\usepackage{xcolor}
\usepackage{xspace}
\usepackage[labelfont=bf]{caption}
\usepackage{soul}
\usepackage{lineno}
\usepackage{amsfonts}
\usepackage{graphicx}
\usepackage{authblk}
\usepackage{xr}
\usepackage[nomessages]{fp}
\usepackage{intcalc}
\usepackage{lipsum}
\usepackage{nameref}
\usepackage{xifthen}
\usepackage{etoolbox}
\usepackage{tikz}
\usetikzlibrary{bayesnet}
\usetikzlibrary{calc} 
\usepackage{graphicx}
\usepackage{amsmath}
\usepackage{booktabs} 
\usepackage{multirow}
\usepackage{bbm}
\usepackage{adjustbox}
\usepackage{url}
\usepackage[acronym, hyperfirst=false]{glossaries}
\usepackage[perpage]{footmisc}
\usepackage{ctable}
\usepackage{tabularx}
\usepackage[position=top]{subcaption}
\usepackage[T1]{fontenc}
\usepackage{titlesec}
\titleformat{\subsubsection}[runin]{\bfseries}{}{}{}[]
\usepackage{lscape}

\newcommand{\todo}[2]{}
\newcommand{\stat}[2]{\scriptsize \textcolor{gray}{#1\textsuperscript{+} #2\textsuperscript{-}}}

\usepackage{verbatim}

\newcommand\ylab{y_i}
\newcommand\xlab{x_i}
\newcommand\predlab{\hat{h}(x_i)}
\newcommand\zlab{z_i}
\newcommand\wlab{w_i}
\newcommand\vlab{v_i}
\newcommand\elab{e_i}
\newcommand{\B}[1]{\boldsymbol{#1}}

\newacronym{phe}{PHE}{Public Health England}
\newacronym{nhs}{NHS}{National Health Service}
\newacronym{sarscov2}{SARS-CoV-2}{severe acute respiratory syndrome coronavirus 2}
\newacronym{uk}{UK}{United Kingdom}
\newacronym{covid}{COVID-19}{coronavirus disease 2019}
\newacronym{pcr}{PCR}{reverse transcription polymerase chain reaction}
\newacronym{lfd}{LFD}{lateral flow device}
\newacronym{lft}{LFT}{lateral flow test}
\newacronym{npi}{NPI}{non-pharmaceutical intervention}
\newacronym{ons}{ONS}{Office for National Statistics}
\newacronym{cis}{CIS}{COVID-19 Infection Survey}
\newacronym{react}{REACT}{REal-time Assessment of Community Transmission}
\newacronym{dag}{DAG}{directed acyclic graph}
\newacronym[longplural={lower tier local authorities}]{ltla}{LTLA}{lower tier local authority}
\newacronym{eb}{EB}{empirical Bayes}
\newacronym{ci}{CI}{credible interval}
\newacronym{voc}{VoC}{variant of concern}
\newacronym{sgtf}{SGTF}{S gene target failure}
\newacronym{sir}{SIR}{Susceptible-Infectious-Recovered}
\newacronym{dtmc}{DTMC}{discrete time Markov chain}
\newacronym{seir}{SEIR}{Susceptible-Exposed-Infectious-Removed}
\newacronym{si}{SI}{Supplementary Information}
\newacronym{ses}{SES}{socioeconomic status}
\newacronym{ode}{ODE}{ordinary differential equation}
\newacronym{lad}{LAD}{local authority district}
\newacronym{la}{LA}{local authority}
\newacronym{delve}{DELVE}{Data Evaluation and Learning for Viral Epidemics}
\newacronym{cmmid}{cmmid}{centre for mathematical modelling of infectious diseases}
\newacronym{mrc}{MRC}{Medical Research Council}
\newacronym{bsu}{BSU}{Biostatistics Unit}
\newacronym{ct}{CT}{chest computed tomography}
\newacronym{cxr}{CXR}{chest radiograph}
\newacronym{covidpos}{COVID$^{+}$\xspace\unskip}{\gls{sarscov2} \gls{pcr}-positive}
\newacronym{covidneg}{COVID$^{-}$\xspace\unskip}{\gls{sarscov2} \gls{pcr}-negative}
\newacronym{cnn}{CNN}{Convolutional Neural Network}
\newacronym{svm}{SVM}{Support Vector Machine}
\newacronym{bnn}{BNN}{Bayesian Neural Network}
\newacronym{ssast}{SSAST}{Self-Supervised Audio Spectrogram Transformer}
\newacronym{mfcc}{MFCCs}{Mel-frequency cepstral coefficients}
\newacronym{uar}{UAR}{Unweighted Average Recall}
\newacronym{roc}{ROC-AUC}{Receiver Operating Charateristic Area Under the Curve}
\newacronym{pr}{PR-AUC}{Precision Recall Area Under the Curve}
\newacronym{mi}{MI}{Mutual Information}
\newacronym{pca}{PCA}{Principal Component Analysis}
\newacronym{pc}{PC}{Principal Component}
\newacronym{fdr}{FDR}{False Discovery Rate}
\newacronym{who}{WHO}{World Health Organization}
\newacronym{abcs}{ABCS}{audio-based COVID-19 screening}
\newacronym{eu}{EU}{expected utility}
\newacronym{tt}{T+T}{Test-and-Trace}
\newacronym{rf}{RF}{random forest}
\newacronym{auc}{AUC}{Area Under the Curve}
\newacronym{gdpr}{GDPR}{General Data Protection Regulation}
\newacronym{tsne}{t-SNE}{t-distributed stochastic neighbor embedding}
\newacronym{ukhsa}{UKHSA}{UK Health Security Agency}

\makeglossaries

\begin{document}









\title{Audio-based AI classifiers show no evidence of improved COVID-19 screening over simple symptoms checkers}

\author[1,2]{Harry Coppock\dag*}
\author[1,3]{George Nicholson*}
\author[1,3,4]{Ivan Kiskin*}
\author[5, 1]{Vasiliki Koutra}
\author[5, 1]{Kieran Baker}
\author[6,7]{Jobie Budd}
\author[8]{Richard Payne}
\author[1]{Emma Karoune}
\author[8]{David Hurley}
\author[8]{Alexander Titcomb}
\author[8]{Sabrina Egglestone}
\author[8,9]{Ana Tendero Ca\~{n}adas}
\author[8]{Lorraine Butler}
\author[1]{Radka Jersakova}
\author[8]{Jonathon Mellor}
\author[8,10]{Selina Patel}
\author[11]{Tracey Thornley}
\author[12]{Peter Diggle}
\author[1]{Sylvia Richardson}
\author[8]{Josef Packham}
\author[1,2,13]{Bj\"orn W.\ Schuller}
\author[5, 1]{Davide Pigoli**}
\author[5, 1]{Steven Gilmour**}
\author[3, 1]{Stephen Roberts**}
\author[1,3]{Chris Holmes}

\affil[1]{The Alan Turing Institute, UK}
\affil[2]{Imperial College London, UK}
\affil[3]{University of Oxford, UK}
\affil[4]{Surrey Institute for People-Centred AI, University of Surrey, UK}
\affil[5]{King's College London, UK}
\affil[6]{Division of Medicine, University College London, UK}
\affil[7]{London Centre for Nanotechnology, University College London, UK}
\affil[8]{UK Health Security Agency, UK}
\affil[9]{School of Applied Sciences, University of Brighton, UK}
\affil[10]{Institute of Health Informatics, University College London, UK}
\affil[11]{University of Nottingham, UK}
\affil[12]{University of Lancaster, UK}
\affil[13]{University of Augsburg, Germany}

\def\thefootnote{\dag}\footnotetext{Corresponding author, \texttt{harry.coppock@imperial.ac.uk}}
\def\thefootnote{*}\footnotetext{Joint first}
\def\thefootnote{**}\footnotetext{Joint senior}
\def\thefootnote{\arabic{footnote}}
\maketitle
\begin{abstract}
    Recent work has reported that respiratory audio-trained AI classifiers
    can accurately predict SARS-CoV-2 infection status. Here, we undertake a large-scale study of audio-based AI classifiers, as part of the UK government's pandemic response. We collect a dataset of audio recordings from 67,842\unskip\  individuals, with linked metadata, of whom 23,514
\unskip\ had positive PCR tests for SARS-CoV-2. 
    In an unadjusted analysis, similar to that in previous works, AI classifiers predict SARS-CoV-2 infection status with high accuracy (ROC-AUC=0.846 [0.838--0.854]). However, after matching on measured confounders, such as self-reported symptoms, performance is much weaker (ROC-AUC=0.619 [0.594--0.644]). Upon quantifying the utility of audio-based classifiers in practical settings, we find them to be outperformed by predictions based on user-reported symptoms.
    We make best-practice recommendations for handling recruitment bias, and for assessing audio-based classifiers by their utility in relevant practical settings. Our work provides novel insights into the value of AI audio analysis and the importance of study design and treatment of confounders in AI-enabled diagnostics.
\end{abstract}
\vspace{.5cm}
Keywords: COVID-19, Machine Learning, Artificial Intelligence, Study Design, Bioacoustics, Causality, Confounding Variables

\vspace{.5cm}



The \gls{covid} pandemic has been estimated by the \gls{who} to have caused 14.9 million excess deaths over the 2020--2021 period (\href{https://www.who.int/news/item/05-05-2022-14.9-million-excess-deaths-were-associated-with-the-covid-19-pandemic-in-2020-and-2021}{\texttt{link}}). 
An accepted public health control measure for emerging infectious diseases is the isolation of infected individuals\cite{Kucharski2020}. As \gls{covid} transmission occurs in both symptomatic and asymptomatic cases\cite{muller2021}, especially prior to nationwide vaccination deployment, a scalable and accurate test for the infection is crucial to avoid general population quarantine. 

This has sparked an intense interest in AI-based classifiers that use respiratory audio data to classify \gls{sarscov2} infection status---here, we refer to this as \textit{\gls{covid} status}---via a digital screening tool that anyone with a smartphone or computer can use\cite{schuller_detecting_2020,laguarta_covid-19_2020,bagad_cough_2020,brown_exploring_2020,imran_ai4covid-19_2020,pinkas_sars-cov-2_2020,hassan_covid-19_2020,han_exploring_2021,chaudhari_virufy_2021,lella_automatic_2021,andreu-perez_generic_2021,coppock_end--end_2021,pahar_covid-19_2021,pizzo_iatos_2021,han_sounds_2022}. In our review, as of July 2022, we found 93 published papers that reported evidence for the potential of audio-based \gls{covid} classification. Of these 93 papers, 75 report an \gls{auc} (or F1) over 0.75 and 44 report performance above 0.90. Table \ref{tab:other-datasets} summarises nine highly cited datasets and corresponding classification performance.


Despite these encouraging results, concerns remain that the prediction models may not be transferable to real-world settings\cite{coppock_end--end_2021, wynants_prediction_2020, roberts_common_2021-1,han_exploring_2021, coppock2021covid, han_sounds_2022}. In some cases data quality may be lowered by, e.g., sampling biases, lack of verification of participants' \gls{covid} status, a long delay between infection and audio recording, or small numbers of \gls{covidpos} individuals\cite{coppock2021covid}. Akin to findings in AI radiographic \gls{covid} detection\cite{degrave2021}, concerns centre around whether the learnt audio features are unique audio biomarkers caused by \gls{covid} in the infected individual, or are due to other confounding signals.


Here, we analyse the largest \gls{pcr}-validated dataset collected to date in the field of \gls{abcs}. We design and specify an analysis plan in advance, to investigate whether using audio-based classifiers can improve the accuracy of \gls{covid} screening over using self-reported symptoms.


Our contribution is as follows:
\begin{itemize}
    \item[] -- We collect a respiratory acoustic dataset of 67,842\unskip\ individuals with linked \gls{pcr} test outcomes, including 23,514
\unskip\ who tested positive for \gls{covid}. To our knowledge, this is the largest \gls{pcr}-validated dataset collected of its kind to date \cite{budd2022}.
    \item[] -- We fit a range of AI classifiers and observe strong \gls{covid} predictive performance (\gls{roc}=0.85), as has been reported in previous studies. However, when controlling for measured confounders by matching, only a small amount of residual predictive variation remains (\gls{roc}=0.62), some of which we attribute to unmeasured confounders. 
    \item[] -- We find the \gls{covid} predictive performance and practical utility of audio-based AI classifiers, as applied in simulated realistic settings, to be no better than classification based upon self-reported symptoms; we replicate this finding by fitting our classifiers in an external dataset.  
    \item[] -- These results suggest that audio-based classifiers learn to predict \gls{covid} via self-reported symptoms and potentially other confounders. Study recruitment based on self-screened symptoms appears to be an important driver of this effect.
    \item[] -- We provide best-practice recommendations on how to address this problem in future studies. 
    \item[] -- Our dataset and codebase is publicly available to enable reproducibility of results and to encourage further research into respiratory audio analysis and bias mitigation in high-dimensional, over-parameterised settings\cite{budd2022}.
\end{itemize}

Our work is timely in highlighting the need for careful construction of machine learning evaluation procedures, aimed at yielding representative performance metrics. The important lessons from this case study on the effects of confounding extend across many applications in AI---where biases are often hard to spot and difficult to control for.
\section*{Results}
\label{sec:results}

\subsection*{Study design}

This study invited volunteers from \gls{react} and \gls{nhs} \gls{tt} to participate between March 2021 and March 2022 on an opt-in basis. Volunteers were directed to the ``Speak up and help beat coronavirus'' web page\cite{speak_up_and_help_beat_url}, where they were instructed to provide audio recordings of four respiratory audio modalities. Demographic and health metadata, along with a validated \gls{pcr} test result, were transferred from  existing \gls{tt}/\gls{react} records. Additional audio-specific metadata were produced from the audio files after collection. The final dataset comprised 23,514
 \gls{covidpos} and 44,328
 \gls{covidneg} individuals. Figure \ref{fig:demographics} summarises the dataset, a more detailed description is provided in Methods, and a full presentation is in the accompanying dataset paper\cite{budd2022}. 

\subsection*{Defining the acoustic target for audio-based \gls{covid} screening}

If a practically effective acoustic signal exists in \gls{sarscov2} infected individuals` respiratory sounds, we propose it would have the following properties:

\begin{itemize}
    \item []\textbf{P1: Caused by \gls{covid}.} \gls{covid} is well known to cause symptoms, such as a new continuous cough, that can be readily self-screened by individuals in the general population. The acoustic target would likewise be linked causally to COVID-19 and it would therefore be more likely to generalise to other contexts and populations than would non-causal associations. 
    \item []\textbf{P2: Not self-identifiable.} The acoustic target would not directly represent \textit{self-identifiable} symptoms 
    that can be self-identified effectively by individuals in the general population. This is because, firstly, it is more straightforward to measure self-identifiable symptoms directly using a questionnaire, rather than measuring them indirectly via audio. Secondly, as we explain below, self-identifiable symptoms can affect enrolment, and may therefore be strongly non-causally associated with \gls{covid} in enrolled subpopulations.
    \item []\textbf{P3: Enables high-utility \gls{covid} screening.} For an audio-based classifier to performed strongly in practical settings, it would possess high sensitivity and specificity, corresponding  to an acoustic signal that would be detectable in a high proportion of \gls{covidpos} individuals and in a low proportion of \gls{covidneg} individuals respectively. We formalise the mathematical relationship linking expected utility, sensitivity and specificity in equation \eqref{eq:expected_utility}.
\end{itemize}

\subsection*{Characterising and controlling recruitment bias}

In audio-based \gls{covid} classification, results can be highly sensitive to the characteristics of the enrolled population. Our study's recruitment protocol is subject to enrolment bias, since the vast majority of individuals in Pillar 2 of the \gls{uk} government's NHS \gls{tt} programme\cite{department_of_health_and_social_care_uk_covid-19_nodate} were \gls{pcr} tested as a direct consequence of reporting symptoms (see Methods). Figures~\ref{fig:demographics}(e) and \ref{fig:demographics}(f) display our participants' symptom profiles, stratified by \gls{covid} infection status. Figure~\ref{fig:bias_intro}(a) presents 
the joint distribution of \gls{covid} status and binary symptoms status as ``Symptoms-based enrolment'', in contrast to Figure~\ref{fig:bias_intro}(b), which presents ``General population enrolment'', based on random sampling from a general population having 2\% \gls{covidpos} prevalence. Note that the dependence between binary symptoms status and \gls{covid} is stronger under Symptoms-based enrolment (population correlation coefficient $\rho$=0.66
\unskip) compared to General population enrolment  ($\rho$=0.15
\unskip).

To clarify the source of enrolment-driven associations, we use a simplified causal model of symptoms-based recruitment (Figure~\ref{fig:causal_model}(a)(i)). 
Enrolment is influenced jointly by \gls{covid} status, self-reported symptoms, and factors such as age and gender (Figure~\ref{fig:knowledge_graph_v1} shows a detailed Bayesian knowledge graph of the recruitment process). Collecting data only from enrolled individuals is, in effect, conditioning on $\elab=1$ at the enrolment node in Figure~\ref{fig:causal_model}(a)(i). Since the enrolment node has directed edges incoming from both \gls{covid} status and self-reported symptoms, conditioning on it induces a non-causal dependence between its parent nodes (in addition to the causal dependence of symptoms on \gls{covid} status). Figure~\ref{fig:causal_model}(a)(ii) displays the moralised undirected graph implied by Figure~\ref{fig:causal_model}(a)(i), conditional on enrolment, with the strong \gls{covid}-to-symptoms dependence represented by a thick line, illustratively labelled $\rho$=0.66
\unskip\ with reference to Figure~\ref{fig:bias_intro}(a). In contrast, Figure~\ref{fig:causal_model}(b)(i) is conditional on random enrolment and does not introduce any additional non-causal association between \gls{covid} status and self-reported symptoms.

If a study's enrolment bias is unaddressed and shared across both training and held-out test sets, a classifier that appears to perform well may not generalise to other datasets\cite{roberts_common_2021, coppock2021covid}. This is due to two effects. First, the classifier may learn to predict using confounding variables that are not causally related to \gls{covid}, but are associated because of their influence on enrolment; e.g., gender, age, or symptoms unrelated to \gls{covid}. Second, even symptoms that are truly causally related to \gls{covid}, such as a new continuous cough, may exhibit \textit{inflated} association with \gls{covid} in the enrolled cohort due to their influence on enrolment.

As well as leading to poor generalisability, audible characteristics that are non-causally but strongly associated with \gls{covid} can obscure any \gls{covid} acoustic signature that may exist. This is illustrated in Figure~\ref{fig:causal_model}(a)(iii), where the association between classifier prediction and \gls{sarscov2} status is mediated by symptoms instead of via a targeted latent COVID acoustic signature. Even in the case of randomised enrolment from the general population, a classifier may learn to predict \gls{sarscov2} status via self-reported symptoms, as opposed to via a latent \gls{covid} acoustic signature, as illustrated in Figure~\ref{fig:causal_model}(b)(iii). 

Here, our goal is to build a classifier whose association with \gls{covid} is mediated by an acoustic signature with the three properties defined above. 
We use established epidemiological methodology known as \textit{matching}\cite{stuart_matching_2010}, whereby study enrolment balances the number of \gls{covidpos} and \gls{covidneg} participants having each combination of potentially audible measured confounding variables. This has the effect of inducing independence between \gls{covid} and these confounders in the matched population, as shown in Figure~\ref{fig:causal_model}(c)(ii). The classifier is then constrained to predict \gls{covid} status either via the latent \gls{covid} acoustic signature, or via unmeasured confounders. 



\subsection*{Primary analyses}
\label{sec:primary_analyses}

\subsubsection*{Pre-specified analysis plan.} We designed and fixed a pre-specified analysis plan, to increase replicability of conclusions\cite{kahan_how_2020}. As part of this advance planning, we detailed the analyses to be conducted and generated the test/validate/train data splits through subsampling of the full dataset. The design of these splits is detailed in Methods, with sample sizes presented in Figure \ref{fig:demographics}(h).

\subsubsection*{Audio-based \gls{covid} prediction performance.}\label{sec:classifier_performance}
Table~\ref{tab:main-results} presents our study's \gls{covid} prediction performance, across nine train/validate/test splits, four modalities, and three models: \gls{ssast}, \glspl{bnn} and openSMILE-\gls{svm}. The \gls{ssast} and \gls{bnn} classifiers consistently outperform the baseline \gls{svm}, and best prediction is achieved with the sentence modality. Reported results are for the \gls{ssast} performance on the sentence modality, unless otherwise stated. Under the Randomised data split the \gls{ssast} classifier achieves high \gls{covid} predictive accuracy of \gls{roc}=0.846 [0.838--0.854]. We hypothesise that this strong predictive accuracy is mainly attributable to enrolment based upon self-reported symptoms, and explore this further in confirmatory analyses below. 

When we control for enrolment bias by matching on age, gender and self-reported symptoms, predictive accuracy drops to a consistently low level of \gls{roc}=0.619 [0.594--0.644] in the Matched test set and \gls{roc}=0.621 [0.605--0.637] in the Longitudinal Matched test set (both trained on the Standard training set). When training instead on our Matched training set, we see a minor improvement in the Matched test set (\gls{roc}=0.635 [0.610--0.660]) and in contrast a slight decrease in prediction accuracy in the Longitudinal Matched test set (\gls{roc}=0.604 [0.587--0.621]). Figure \ref{fig:illustrative} illustrates these different experimental settings and the corresponding classification performance. A cluster analysis is also performed on the \gls{ssast} learnt representations, detailed in Supplementary Note 2, visually demonstrating the effect of decoupling measured confounders and \gls{covid} status. To explore whether classifier performance might be higher in some matched groups than in others we calculated \gls{roc} within matched strata (Figure~\ref{fig:stratified_roc-auc}), observing the estimates and \glspl{ci} to be consistent with a homogenously low predictive score of \gls{roc}=0.62 across strata.  

\subsection*{Confirmatory analyses and validation}

\subsubsection*{The additional predictive value of \gls{abcs}.} 
Audio-based classifiers can be useful in practice if they deliver improved performance relative to classifiers based on self-identifiable symptoms. Moreover, it is beneficial to assess the performance of \gls{abcs} classifiers in test sets that reflect the application of the testing protocol in a real-life setting. 
Here, we generate a ``general population'' test set, through balanced sub-sampling without replacement from our combined Standard and Longitudinal test sets, so as to capture the age/gender/symptoms/\gls{covid} profile of the general population during the pandemic. 
Specifically, the proportion of symptomatic individuals is set at 65\% in the \gls{covidpos} subgroup\cite{sah_asymptomatic_2021}, compared to a setting of one of (10\%, 20\%, 30\%) symptomatic individuals in the \gls{covidneg} subgroup; the age distribution is constrained to be the same in both \gls{covidpos} and \gls{covidneg} subgroups; and with males/females balanced in 1:1 ratio in each \gls{covidpos}/\gls{covidneg} subgroup. 
We benchmark the \gls{covid} predictive performance of the audio-based \gls{ssast} classifier against the  performance attainable through \gls{rf} classifiers trained on self-identifiable symptoms and demographic data (a ``Symptoms'' \gls{rf} classifier). We also include in the benchmarking an \gls{rf} classifier taking as inputs the audio-based \gls{ssast} probabilistic outputs alongside self-identifiable symptoms and demographic data (``Symptoms+Audio'' \gls{rf} classifier). Training for all three classifiers is performed in our Standard training set. 
The resulting ROC curves are shown in Figure~\ref{fig:audio_vs_symptoms_genpop}(a)-(c). Focusing on the general population with 20\% of \gls{covidneg} symptomatic in Figure~\ref{fig:audio_vs_symptoms_genpop}(b), the combined Symptoms+Audio \gls{rf} classifier 
(\gls{roc}=0.787 [0.772--0.801], 95\% DeLong CI)  
offers a significant 
(p=$9.7\times 10^{-11}$, DeLong test) 
but small increase in predictive accuracy over the Symptoms \gls{rf} classifier  
(\gls{roc}=0.757 [0.743--0.771], 95\% DeLong CI)
, which in turn yields a less significant increase in \gls{roc} compared to the Audio-only classifier
(p=0.0033) 
(\gls{roc}=0.733 [0.717−0.748], 95\% DeLong CI). 

We replicate these findings using an external dataset\cite{han_sounds_2022}, observing qualitatively similar results with a Symptoms classifier (\gls{roc}=0.79 [0.71--0.87]) outperforming an \gls{ssast} audio-based classifier (\gls{roc}=0.68 [0.59--0.77]) in a general population test set simulated from the external test set\cite{han_sounds_2022}, with 20\% of \gls{covidneg} individuals symptomatic. We observe similar results when comparing a Symptoms classifier to our \gls{ssast} and Han \textit{et al.}'s \gls{cnn} directly on the external test set\cite{han_sounds_2022}: Symptoms classifier \gls{roc}=0.81 [0.76--0.86]; \gls{ssast} audio-based classifier \gls{roc}=0.68 [0.62--0.74]; CNN audio-based classifier\cite{han_sounds_2022} \gls{roc}=0.66 [0.60--0.71] (see also Figure~\ref{fig:cambridge_raw_test_set_results}). Reported results for \gls{ssast} and  Han \textit{et al.}'s \gls{cnn} are for the `cough' modality however, we see similar results for both `breathing' and `voice'.


\subsubsection*{Translating prediction accuracy into utility.} To characterise the practical benefit of  \gls{abcs} in any particular testing setting, we can specify the utility $u_{\hat{y}, y}$ of predicting $\hat{y}\in \{0,1\}$ for a random individual, in the targeted testing population, whose true COVID status is $y\in \{0,1\}$, and calculate the per-test \gls{eu} as
\begin{align}
\nonumber
    \text{EU}\equiv \mathbb{E}[\text{utility} \mid \boldsymbol{u}, \pi, \text{sensitivity}, \text{specificity}] &= \pi\left[ (u_{1,1} - u_{0,1}) \times \text{sensitivity} + u_{0,1}\right]\\
    \label{eq:expected_utility}
    & \hspace{.25cm} + (1 - \pi)\left[ (u_{0,0} - u_{1,0}) \times \text{specificity} + u_{1,0}\right],
\end{align}
where $\pi$ is the \gls{covid} prevalence in the tested population (equation \eqref{eq:expected_utility} is derived in Methods). The \gls{eu} is increasing in both sensitivity and specificity, with their relative weights depending on prevalence $\pi$ and utility $\B{u}$.
Note that it is not only $\pi$ and $\B{u}$ that are context-dependent: the sensitivity and specificity of any particular \gls{covid}-classifier depends on the characteristics of the targeted testing population, as illustrated by the effects of variation in the proportion of \gls{covidneg} individuals that are symptomatic across Figure~\ref{fig:audio_vs_symptoms_genpop}(a)-(c).

Consider the following illustrative utility function, measured in units of the number of infections prevented:
\begin{align*}
    u_{1,1}&=R_t - \varepsilon&[\text{True positive result for \gls{covidpos}, $R_t$ infections prevented on average}]\\
    u_{1,0}&=-\varepsilon&[\text{False positive for \gls{covidneg}, $-\varepsilon$ is negative impact of self-isolation}]\\
    u_{0,0}&=0&[\text{True negative for \gls{covidneg}}]\\
    u_{0,1}&=-\delta &[\text{False negative for \gls{covidpos}, causing $\delta$ additional infections on average}]
\end{align*}
In the above, there are three specified parameters: (i) the number of cases prevented by intervention on a single individual is specified as the effective reproduction number, $R_t\geq 0$, i.e., the average number of infections that person would cause under no intervention, assuming that all individuals with a positive result follow self-isolation guidance and cause no transmission; (ii) $\varepsilon\geq 0$ measures the cost of intervention (e.g., the negative impact on health or education resulting from self-isolation); and (iii) $\delta\geq 0$ is the expected number of additional infections caused by a false-negative result (e.g., due to reduced caution and increased social mixing following a negative result).

Figure~\ref{fig:audio_vs_symptoms_genpop}(d)-(f) shows maximum EU, as a function of prevalence, for settings of $R_t \in \{1,1.5\}$ and $\varepsilon \in \{0.02,0.2\}$ and with $\delta=0$ (corresponding results for $\delta=0.25$ are shown Figure~\ref{fig:audio_vs_symptoms_genpop_delta_0.25}). The maximisation is performed point-wise with respect to sensitivity and specificity across the corresponding ROC curves in Figure~\ref{fig:audio_vs_symptoms_genpop}(a)-(c)). The utility of all classifiers decreases as the percentage of \gls{covidneg} symptomatics increases from 10\% to 30\% in Figure~\ref{fig:audio_vs_symptoms_genpop}(d)-(f), the intuition being that it is more difficult to distinguish \gls{covidneg} from the 65\% symptomatic \gls{covidpos} population. When we compare the Symptoms+Audio RF classifier to the Symptoms RF classifier, neither is generally optimal, with each classifier showing greater EU than the others for some values of $(\pi, R_t, \varepsilon)$.

\subsubsection*{Exploratory approaches to identify the influence of unmeasured confounders.}
\label{sec:unaccounted}
We explore whether the residual \gls{covid} prediction performance of \gls{roc}=0.62, in the matched test set, is truly attributable to the targeted acoustic signature, or whether it stems from unmeasured confounders, such as the audible recording environment or unreported symptoms. We propose two complementary exploratory methods which are presented in greater detail in Supplementary Note 1. 

Method 1 investigates how much of the residual predictive variation persists when we map all Matched test set samples to the the first $k$ Principal Components (PCs) of the \gls{covidneg} samples. We train a classifier on \gls{covid} detection in this $k$ dimensional space and hypothesise that below a threshold value for $k\leq \tau$, correct classification is due to confounding in the signal. The value of $\tau$ is determined by running a calibration experiment and is set to 14 for the sentence modality. By removing these correctly classified cases to form a curated Matched test set,  we see a drop in \gls{ssast} performance to \gls{uar}=0.51. 

Method 2 examines how much residual predictive variation persists when we map each \gls{covidpos} participant to their nearest neighbour (NN) \gls{covidneg}, using a robust distance metric in openSMILE space. Predictive variation that persists in the space spanned by \gls{covidneg} individuals is then attributed to unmeasured confounders. After the \gls{covidpos}-to-\gls{covidneg} NN mapping, \gls{svm}, Matched test set \gls{roc} drops from 0.60 to 0.55. We interpret this persistent component of predictive variation after the mapping to \gls{covidneg} individuals (i.e., that \gls{roc} drops only to 0.55 as opposed to 0.50) as pointing to some degree of unmeasured confounder contributing to the score of \gls{roc}=0.60 in the Matched test set.

\section*{Discussion}
\label{sec:discussion}


\gls{covid} is well known to be causally related to particular self-identifiable symptoms, such as a new continuous cough. This has allowed such symptoms to be used by governments during the pandemic as a basis for population intervention to control disease spread, e.g., a triage tool for individuals, via self-screening and without recourse to audio recording. It is therefore desirable to develop audio-based classifiers that can augment and complement the information provided by self-identifiable \gls{covid}-specific symptoms, i.e., to learn clinically valuable \textit{latent} acoustic signatures caused by \gls{covid}.   

Problematically, enrolment based on symptoms has the potential not only to artificially inflate the association between \gls{covid} and its particular symptoms, but also to introduce association between \gls{covid} and symptoms that are not \gls{covid} specific. Furthermore, enrolment based upon other characteristics, such as gender and age, may additionally introduce non-causal \gls{covid}-to-gender or \gls{covid}-to-age associations in the enrolled subpopulation, \textit{possibly interacting with symptoms}. 

Under such recruitment bias, classifiers trained to predict \gls{covid} in enrolled subpopulations may learn to predict \textit{self-identifiable} \gls{covid}-specific symptoms, thereby providing no additional utility beyond a classifier trained directly on those self-screened symptoms. Worse, the classifier may learn to predict age/gender/non-\gls{covid}-specific symptoms as proxies for \gls{covid} in the enrolled subpopulation, in which case its performance will not generalise to subpopulations unaffected by the same recruitment bias. 

Han \textit{et al.} \cite{han_sounds_2022} examine several aspects of recruitment bias in \gls{abcs}, simulating the effects of biases introduced by age, gender, language, and by the same individuals appearing in both train and test sets. While their dataset is approximately balanced with respect to age and gender (across \gls{covidpos} and \gls{covidneg} subgroups), it is imbalanced with respect to self-reported symptoms (84\% of their \gls{covidpos} are symptomatic, compared to 49\% of \gls{covidneg}, ref.\ \cite[Fig 1e]{han_sounds_2022}). Such imbalance with respect to symptoms is also present in our study (prior to matching) and other studies for which data, including symptoms, are available (see Figure \ref{fig:dataset_symp}).

We make the following recommendations aimed at clarifying the effects of recruitment bias and mediation by self-identifiable symptoms in future studies:
\begin{enumerate}
        \item \textbf{Collect and disseminate metadata.} Repositories of audio samples should include details of the study recruitment criteria and relevant metadata (e.g., gender, age, symptoms, location, time since \gls{covid} test), so that data can be filtered for quality and for relevance to hypothesis, and so that bias from measured confounders can be characterised and controlled if necessary.
        \item \textbf{Characterise and control recruitment bias.} Analyse data using methods that acknowledge and control for the effects of recruitment bias. We approached this by matching on measured confounders in our test and/or training sets.
        \item \textbf{Design studies with bias control in mind.} Matching leads to reduced sample size when performed post-recruitment, so it can be beneficial to design observational studies that recruit participants to maximise the potential for matching on measured confounders in the enrolled cohort. 
        \item \textbf{Focus on the added predictive value of classifiers.} Quantify the \textit{additional} predictive value offered by classifiers as compared to standard methods. 
        \item \textbf{Assess classifier performance in targeted settings.} Measures of classifiers' predictive accuracy, such as \gls{roc}, sensitivity, and specificity vary depending upon the characteristics of the targeted population (e.g., according to prevalence and the proportion of \gls{covidpos} and \gls{covidneg} individuals that are symptomatic). Where possible, apply the classifier in a test set that reflects the appropriate application setting, e.g., by subsampling a test set representative of the general population, as we do here. 
        \item \textbf{Examine classifiers' expected utility.}  We can specify utilities for each testing outcome, i.e., quantify the average benefit accrued from a true positive, the different benefit of a true negative, and similarly the costs attached to false positives and false negatives. Then the expected utility provides a highly context-specific score for 
        quantifying and comparing classifiers' performance.
       \item \textbf{Out-of-study replication.} Replication studies could be performed in randomly sampled cohorts, or in pilot studies in real-world prediction settings with domain-specific utility functions. There are extra challenges when performing out-of-study replication; in particular, the audio-capture protocols might differ. It would facilitate replication if standardised protocols for audio data gathering are collaboratively developed. 
\end{enumerate}

We conclude by outlining some limitations of our study, dataset, and findings. There are potentially subtle unmeasured confounders across our recruitment channels \gls{react} and \gls{tt}. For example, \gls{pcr} testing in \gls{tt} usually occurs in the days after self-screening of symptoms whilst, in \gls{react}, \gls{pcr} tests are more likely to occur on a date approximately pre-determined by study researchers, and so be independent of participants' symptoms. We attempted to control for such unmeasured confounders by including recruitment channel as one of the matched variables in the test set. Despite matching on measured confounders, some residual predictive variation persists (\gls{roc}=0.62). Our exploratory approaches for characterising this residual predictive variation (Methods and Supplementary Note 1) suggest that some of this residual performance may be due to unmeasured confounders, but these results are not conclusive. Our results are suggestive of little utility of \gls{abcs} in practice relative to symptoms-based screening (Figure~\ref{fig:audio_vs_symptoms_genpop}). The development of more sophisticated methods for training audio-based models in the presence of symptoms data and recruitment bias is a worthwhile and active area of research which, alongside careful design and replication of studies, will eventually provide full clarity on the potential of \gls{abcs} as a tool to protect public health.




\section*{Methods}
\label{sec:methods}

\subsection*{Dataset and study design}

This section contains an overview of how the dataset was collected, its characteristics, and its underlying study design. More in-depth descriptions are provided in two accompanying papers, in which 
Budd et al.\cite{budd2022} report a detailed description of the full dataset, and Pigoli et al.\cite{Pigoli2022} present the rationale for and full details of the statistical design of our study. 

\paragraph{Recruitment Sources.} Our main sources of recruitment were: the \gls{react} study and the \gls{nhs} \gls{tt} system. \gls{react} is a prevalence survey of \gls{sarscov2} based on repeated cross-sectional samples from a representative subpopulation defined via (stratified) random sampling from England's \gls{nhs} patient register \cite{chadeau-hyam_react-1_2021}. The \gls{nhs} \gls{tt} service was a key part of the UK government's \gls{covid} recovery strategy for England, ensuring that anyone developing \gls{covid} symptoms could be swab tested, followed by the tracing of recent close contacts of any individuals testing positive for \gls{sarscov2} (ref. \cite{department_of_health_and_social_care_uk_covid-19_nodate}).

\paragraph{Criteria for enrolment.} Enrolment for both the \gls{react} and \gls{nhs} \gls{tt} recruitment channels was performed on an opt-in basis. Individuals participating in the \gls{react} study were presented with the option to volunteer for this study. For the \gls{nhs} \gls{tt} recruitment channel, individuals receiving a \gls{pcr} test from the \gls{nhs} \gls{tt} Pillar 2 scheme were invited to take part in research (Pillar 1 tests refer to \textit{``all swab tests performed in \gls{phe} labs and \gls{nhs} hospitals for those with a clinical need, and health and care workers''}, and Pillar 2 comprises \textit{``swab testing for the wider population''} \cite{department_of_health_and_social_care_uk_covid-19_nodate}). 
The guidance provided to potential participants was that they should be at least 18 years old, have taken a recent swab test (initially no more than 48 hours, changing to 72 hours on 2021-05-14), agree to our data privacy statement, and have their \gls{pcr} barcode identifier available which was then internally validated. 

\paragraph{Audio recordings.} Participants were directed to the ``Speak up and help beat coronavirus'' web page. Here, after agreeing to the privacy statement and completing the survey questions, participants were asked to record four audio clips. The first involved the participant reading out a sentence of text, ``I love nothing more than an afternoon cream tea''. This sentence was designed to contain a range of different vowel and nasal sounds. This was followed by three successive sharp exhalations, taking the form of a `ha' sound. The final two recordings involved the participant performing volitional/forced coughs, once, and then three times in succession. Recordings were saved in \texttt{.wav} format. Smart phones, tablets, laptops and desktops were all permitted. The audio recording protocol was homogenised across platforms to reduce the risk of bias due to device types.

\paragraph{Demographic and clinical/health metadata.} Existing metadata such as age, gender, ethnicity and location was transferred from linked \gls{tt}/\gls{react} records. Participants were not asked to repeat this information to avoid survey fatigue. An additional set of attributes, hypothesised to pose the most utility for evaluating the possibility for \gls{covid} detection from audio, was collected in the digital survey. This was in line with \gls{gdpr} requirements that only the personal data necessary to the task should be collected and processed. This set included current displaying symptoms, the full set of which are detailed in Figure~\ref{fig:demographics}(e)-(f), and long term respiratory conditions, such as Asthma. To control for different dialects/accents, to complement location and ethnicity, participants' first language was also collected. 
Finally, the test centre where the \gls{pcr} was conducted was recorded. This enabled the removal of submissions when cases were linked to faulty test centre results. A full set of the dataset attributes can be found in Budd \textit{et al.} \cite{budd2022}

\paragraph{Final dataset.} 
The final dataset is downstream of a quality control filter (see Figure \ref{fig:demographics} (g)), in which a total of 5,157 records were removed, each with one or more of the following characteristics (a) missing response data (missing a PCR test); (b) missing predictor data (any missing audio files or missing demographic/symptoms metadata); (c) audio submission delays exceeding 10 days post test result; (d) self-inconsistent symptoms data; (e) PCR testing lab under investigation for unreliable results; (f) participant age under 18 (g) sensitive personal information detected in audio signal (see \cite[Figure 3(d)]{budd2022}). Pigoli et al.\cite{Pigoli2022} presents these implemented filters in full and the rationale behind each one. The final collected dataset, after data filtration, comprised 23,514
 \gls{covidpos} and 44,328
 \gls{covidneg} individuals recruited between March 2021 and March 2022. Please note that the sample size here differs to that in our accompanying papers, in which Budd et al.\cite{budd2022} reported numbers before the data quality filter was applied, while our statistical study design considerations, detailed in Pigoli et al.\cite{Pigoli2022}, focused on data from the restricted date range spanning March 2021 to November 2021. We note the step-like profile of the \gls{covidneg} count is due to the six \gls{react} rounds where a higher proportion of \gls{covidneg} participants were recruited compared to the \gls{tt} channel. As detailed in the geo-plots in Figure~\ref{fig:demographics}a-b, the dataset achieves a good coverage across England, with some areas yielding more recruited individuals than others. We are pleased to see no major correlation between geographical location and \gls{covid} status, shown graphically in Figure~\ref{fig:demographics}c, with Cornwall displaying the highest level of \gls{covid} imbalance with 0.8\% difference in \% proportion of \gls{covidpos} and \gls{covidneg} cases.

\subsubsection*{Data splits.}
In our pre-specified analysis plan we defined three training sets and five test sets so as to define a range of analyses in which we investigate, characterise, and control for the effects of enrolment bias in our data:
\begin{itemize}
    \item [] \textbf{`Randomised' train and test set.} A participant-disjoint train and test set was randomly created from the whole dataset, similar to methods in previous works.
    \item [] \textbf{`Standard' train and test set.} Designed to be a challenging, out-of-distribution evaluation procedure. Carefully selected attributes such as geographical location, ethnicity and first language are held out for the test set. The Standard test set was also engineered to over represent sparse combination of categories such as older \gls{covidpos} participants\cite{Pigoli2022}.
    \item [] \textbf{`Matched' train and test sets.} Numbers of \gls{covidneg} and \gls{covidpos} participants are balanced within each of several key strata. Each stratum is defined by a unique combination of measured confounders, including binned age, gender and a number of binary symptoms (e.g., cough, sore throat, shortness of breath; see Methods for full description). 
    \item [] \textbf{`Longitudinal' test set.} To examine how classifiers generalised out-of-sample over time, the Longitudinal test set was constructed only from participants joining the study after 29\textsuperscript{th} November 2021.
    \item [] \textbf{`Matched Longitudinal' test set.} Within the Longitudinal test set, Numbers of \gls{covidneg} and \gls{covidpos} participants are balanced within each of several key strata, similarly as in the Matched test set above. 
\end{itemize}
The supports for each of these splits are detailed in Figure \ref{fig:demographics}(g).

\subsection*{Machine Learning Models}

For the task of \gls{covid} detection from audio, three separate models were implemented, each representing an independent machine learning pipeline. Collectively, these three models span the machine learning research space well, ranging from established baseline to current state of the art in audio classification technologies, and are visually represented in Figure \ref{fig:models}. Additionally, we fitted an \gls{rf} classifier to predict \gls{covid} status from self-reported symptoms and demographic data.  The outcome used to train and test each of the prediction models was a participant's \gls{sarscov2} \gls{pcr} test result. Each model's inputs and predictors and how they are handled are detailed in dedicated sections below. Wherever applicable we have reported our study's findings in accordance with TRIPOD statement guidelines\cite{collins_transparent_2015}. The following measures were used to assess model performance: area under the receiver operating characteristic curve (ROC-AUC), area under the precision-recall curve (PR-AUC), and unweighted average recall (UAR; also known as balanced accuracy).  Confidence intervals for ROC-AUC, PR-AUC and UAR are based on the normal approximation method\cite{hanley_mcneil_1982}, unless otherwise stated to be calculated by the DeLong method\cite{delong_comparing_1988}. 

\paragraph*{openSMILE-\gls{svm}.} For our baseline model we defaulted to the widely used openSMILE-\gls{svm} approach \cite{Eyben10-OTM}. Here, 6,373 hand crafted features (the ComParE 2016 set) such as zero-crossing rate and shimmer, which have been shown to represent human paralinguistics well, are extracted from the raw audio form. These features are then concatenated to form a 6,373 dimensional vector, {$f_{openSMILE}(\mathbf{w}) \to \mathbf{v}$ where the raw wave form, $\mathbf{w} \in \mathbb{R} ^{n}$, $n =$ clip duration in seconds * sample rate and $\mathbf{v} \in \mathbb{R}^{6,373}$. $\mathbf{v}$ is then normalised prior to training and inference. A linear \gls{svm} is fitted to this space and tasked with binary classification. We select the optimal \gls{svm} configuration based on the validation set before then retraining on the combined train-validation set.

\paragraph*{ResNet-50 \gls{bnn}.} 
\glspl{bnn} provide estimates of uncertainty, alongside strong supervised classification performance, which is desirable for real-world use cases, especially involving clinical use. \glspl{bnn} are naturally suited to Bayesian decision theory, which benefits decision-making applications with different costs on error types (e.g., assigning unequal weighting to errors in different \gls{covid} outcome classification) \cite{vadera2021post,cobb2018loss}. We thus supply a ResNet-50\cite{he_deep_2016} \gls{bnn} model. The base ResNet-50 model showed initial strong promise for \gls{abcs} \cite{laguarta_covid-19_2020}, further motivating its inclusion in this comparison. We achieve estimates of uncertainty through Monte-Carlo Dropout to achieve approximate Bayesian inference over the posterior as in \cite{galdropout}. We opt to use the pre-trained model for a warm start to the weight approximations, and allow full re-training of layers.

The features used to create an intermediate representation, as input to the convolutional layers, are Mel filterbank features with default configuration from the VGGish \href{https://github.com/tensorflow/models/blob/master/research/audioset/vggish/vggish_input.py}{GitHub}: ${\mathbf{X}}_i \in {\mathbb{R}}^{96\times 64}$, 64 log-mel spectrogram coefficients using 96 feature frames of 10\,ms duration, taken from a re-sampled signal at 16\,kHz. Each input signal was divided into these two-dimensional windows. For evaluation, the mean prediction over feature windows was taken per audio recording, to produce a single decision per participant. To make use of the available uncertainty metrics, Supplementary Note 3 details an uncertainty analysis over all audio modalities for a range of train-test partitions.

\paragraph*{\gls{ssast}.} In recent years transformers \cite{vaswani_attention} have started to perform well in high-dimensional settings such as audio \cite{baevski_wav2vec2, dosovitskiy_vit}. This is particularly the case when models are first trained in a self-supervised manner on unlabelled audio data. We adopt the \gls{ssast} \cite{Gong_ssast} which is on a par with current state of the art for audio event classification. Raw audio is first resampled to 16\,kHz and normalised before being transformed into Mel filter banks. Strided convolutional neural layers are used to project the Mel filter bank to a series of patch level representations. During self-supervised pretraining random patches are masked before all patches are passed to a transformer encoder. The model is trained to jointly reconstruct the masked audio and to classify the order at which the masked audio occurs. The transformer is made up of 12 multihead attention blocks. The model is trained end to end, with gradients being passed all the way back to the convolutional feature extractors. The model is pre-trained on a combined set of AudioSet-2M \cite{Gemmeke2017} and Librispeech \cite{panayotov2015}, representing over 2 million audio clips for a total of 10 epochs. The model is then fine-tuned in a supervised manner on the task of \gls{covid} detection from audio. Silent sections of audio recordings are removed before then being resampled to 16\,kHz and normalised. Clips are cut/zero-padded to a fixed length of 5.12\,s, corresponding to approximately the mean length of audio clip. The output representations are mean pooled before being fed through a linear projection head. No layers are frozen and again the model is trained end-to-end. The model is fine-tuned for a total of 20 epochs. The model is evaluated on the validation set at the end of each epoch and its weigths are saved. At the end of training the best performing model, over all epochs, is chosen.

\paragraph*{\gls{rf} classifier (self-reported symptoms and demographic data as inputs).} In order to predict \gls{sarscov2} infection status from self-reported symptoms and demographic data, we applied an \gls{rf} classifier with default settings. 
In our dataset, predictor variables for the Symptoms \gls{rf} classifier on our dataset comprised: Cough, Sore throat, Asthma, Shortness of breath, Runny/blocked nose, A new continuous cough, COPD or emphysema, Other respiratory condition, Age, Gender, Smoker status, and Ethnicity. 
In Han et al.'s dataset predictor variables for the Symptoms \gls{rf} classifier comprised: tightness, dry cough, wet cough, runny, chills, smell/taste loss, muscle ache, headache, sore throat, short breath, dizziness, fever, runny blocked nose, Age, Gender, Smoker status, Language and Location.
Prior to training, categorical attributes were one-hot encoded. No hyperparameter tuning was performed, and models were trained on the combined Standard train and validation sets. For the hybrid Symptoms+Audio \gls{rf} classifier, the outputted predicted \gls{covidpos} probability from an audio-trained \gls{ssast} is appended as an additional input variable to the self-reported symptoms and demographic variables listed above.

\subsection*{Matching methodology.} 

The Matched test set was constructed by exactly balancing the numbers of COVID+ and COVID- individuals in each stratum where, to be in the same stratum, individuals must be matched on all of (recruitment channel) x (10-year wide age bins) x (gender) x (all of six binary symptoms covariates). The six binary symptoms matched on in the Matched test set were: Cough, Sore throat, Asthma, Shortness of breath, Runny/blocked nose, and ``At least one symptom''. The resulting Matched test set comprised 907 COVID+ and 907 COVID- participants. The Matched training set was constructed similarly to the Matched test set, though with slightly different strata, so as to increase available sample size. For the Matched training set individuals were matched on all of:  (10-year wide age bins) x (gender) x (all of seven binary covariates). The seven binary covariates used for the Matched training set were: Cough, Sore throat, Asthma, Shortness of breath,Runny/blocked nose, COPD or Emphysema, and Smoker status. The resulting Matched training set comprised 2,599 COVID+ and 2,599 COVID- participants. 


\subsection*{Quantifying the expected utility of a testing protocol.} 
We consider the action of applying a particular testing protocol to an individual randomly selected from a population. The four possible outcomes $O_{\hat{y},y}$ are
\begin{align}
    O_{\hat{y},y} &:= \text{[Predict Covid status as $\hat{y}$]} \text{  AND  } \text{[True Covid status is $y$]}
\end{align}
for predicted Covid status $\hat{y} \in \{0,1\}$ and true Covid status $y \in \{0,1\}$. We denote the probabilities of outcome $O_{\hat{y},y}$ by
\begin{align}
    p_{\hat{y},y}&:= \mathbb{P}(O_{\hat{y},y})
\end{align}
and use $u_{\hat{y},y}$ to denote the combined utility of the consequences of outcome $O_{\hat{y},y}$. For a particular population prevalence proportion, $\pi$, the $p_{\hat{y},y}$ are subject to the  constraints
\begin{align}
\label{eq:deriving_utility_prev_1}
    p_{0,1} + p_{1,1}&= \pi\\
\label{eq:deriving_utility_prev_2}
    p_{0,0} + p_{1,0}&= 1 - \pi\ ,
\end{align}
leading to the following relationships, valid for $\pi \in (0,1)$, involving the sensitivity and specificity of the testing protocol:
\begin{align}
\label{eq:deriving_utility_sens}
    \text{sensitivity} &\equiv \frac{p_{1,1}}{p_{1,1} + p_{0,1}} = \frac{p_{1,1}}{\pi}\\
\label{eq:deriving_utility_spec}
    \text{specificity} &\equiv \frac{p_{0,0}}{p_{0,0} + p_{1,0}} = \frac{p_{0,0}}{1 - \pi} .
\end{align}
The expected utility is:
\begin{align}
\label{eq:deriving_utility_0}
    \text{EU} &= \sum_{\hat{y} \in \{0,1\}} \sum_{y \in \{0,1\}} u_{\hat{y},y} p_{\hat{y},y}\\ 
\label{eq:deriving_utility_1}
    &= u_{1,1} p_{1,1} + u_{0,1} (\pi - p_{1,1}) + u_{0,0} p_{0,0} + u_{1,0} (1 - \pi - p_{0,0})\\
\label{eq:deriving_utility_2}
    &= \pi[(u_{1,1} - u_{0,1}) \times \text{sensitivity} + u_{0,1}] + (1-\pi) [(u_{0,0} - u_{1,0})\times \text{specificity} + u_{1,0}]\ ,
\end{align}
where \eqref{eq:deriving_utility_prev_1} and \eqref{eq:deriving_utility_prev_2} are substituted in \eqref{eq:deriving_utility_0} to obtain \eqref{eq:deriving_utility_1}, and \eqref{eq:deriving_utility_sens} and \eqref{eq:deriving_utility_spec} are substituted in \eqref{eq:deriving_utility_1} to obtain \eqref{eq:deriving_utility_2}.

\section*{Ethics} This study has been approved by The National Statistician’s Data Ethics Advisory Committee (reference NSDEC(21)01) and the Cambridge South NHS Research Ethics Committee (reference 21/EE/0036) and Nottingham NHS Research Ethics Committee (reference 21/EM/0067). All participants reviewed the provided participant information and gave their informed consent to take part in the study.

\section*{Data and Code Availability}

\paragraph*{Data} 

To obtain access to this dataset, named `The UK COVID-19 Vocal Audio Dataset', interested parties may submit their requests to UKHSA at \texttt{DataAccess@ukhsa.gov.uk}. Access is subject to approval and completion of a data sharing contract. For information on how one can apply for UKHSA data, please visit: \href{https://www.gov.uk/government/publications/accessing-ukhsa-protected-data/accessing-ukhsa-protected-data}{https://www.gov.uk/government/publications/accessing-ukhsa-protected-data/accessing-ukhsa-protected-data}. Audio data are provided in \texttt{.wav} format, with four files (one for each recording) for each of the 72,999 participants (unless missing). Metadata are provided in three \texttt{.csv} files, linked by a participant identifier code.

Although the dataset is fully anonymised, and therefore does not contain any personal data, it has been deposited as safeguarded data in line with the privacy notice provided to participants. Safeguarded data can be used for non-commercial, commercial and teaching projects.


 \paragraph*{Code} The code-base developed for this project can be found at this public GitHub repository: \href{https://github.com/alan-turing-institute/Turing-RSS-Health-Data-Lab-Biomedical-Acoustic-Markers}{https://github \\ .com/alan-turing-institute/Turing-RSS-Health-Data-Lab-Biomedical-Acoustic-Markers}. Here, instructions to replicate our experimental environment and run our experiments are provided.
\clearpage

\section*{Tables}
\def\arraystretch{1.2}
\begin{table}[!ht]
    \centering
    \begin{tabularx}{\textwidth}{XXX||XXXX||XXXX||X}
    \specialrule{.2em}{.1em}{.1em}
       \multicolumn{3}{c||} {Train } & \multicolumn{4}{c||} {Standard \stat{9379}{16518}} & \multicolumn{4}{c||} {Match \stat{2599}{2599}} & Random \stat{20000}{37665}\\ \cline{2-12}
        \multicolumn{3}{c||} {Test} & Standard \stat{3820}{7301} & Match \stat{907}{907} & Long \stat{10315}{20509} & Long Match \stat{2098}{2098} & Standard \stat{3820}{7301} & Match \stat{907}{907} & Long \stat{10315}{20509} & Long Match \stat{2098}{2098} & Random \stat{3514}{6663}\\ \specialrule{.125em}{.1em}{.1em}
        \multirow{9}{0pt} {Sent ence} & \multirow{3}{*} {SVM} & UAR & 0.669 & 0.566 & 0.699 & 0.570 & 0.658 & 0.567 & 0.646 & 0.579 & 0.721 \\ \cline{3-12}
        &  & ROC & 0.732 & 0.596 & 0.766 & 0.591 & 0.714 & 0.600 & 0.693 & 0.597 & 0.796 \\ \cline{3-12}
         &  & PR & 0.578 & 0.574 & 0.625 & 0.580 & 0.553 & 0.583 & 0.515 & 0.576 & 0.686 \\ \cline{2-12}
         & \multirow{3}{0pt}{SSAST} & UAR & \textbf{0.733} & \textbf{0.594} & \textbf{0.739} & \textbf{0.583} & 0.692 & 0.602 & 0.666 & 0.572 & \textbf{0.763} \\ \cline{3-12}
         &  & ROC & \textbf{0.800} & 0.619 & \textbf{0.818} & \textbf{0.621} & 0.760 & \textbf{0.635} & 0.732 & 0.604 & \textbf{0.846} \\ \cline{3-12}
         & & PR & \textbf{0.684} & 0.594 & \textbf{0.715} & \textbf{0.594} & 0.631 & 0.626 & 0.590 & 0.579 & \textbf{0.774} \\ \cline{2-12}
         & \multirow{3}{0pt}{BNN} & UAR & 0.685 & 0.586 & 0.702 & 0.566 & 0.703 & \textbf{0.604} & 0.687 & 0.581 & 0.702 \\ \cline{3-12}
         &  & ROC & 0.776 & \textbf{0.623} & 0.804 & 0.614 & 0.767 & 0.634 & 0.749 & 0.610 & 0.834 \\ \cline{3-12}
         & & PR & 0.645 & \textbf{0.613} & 0.689 & 0.593 & 0.634 & \textbf{0.629} & 0.619 & 0.593 & 0.752 \\ 
         \specialrule{.125em}{.1em}{.1em}
        \multirow{9}{0pt} {Three coughs} &  \multirow{3}{0pt} {SVM} & UAR & 0.669 & 0.555 & 0.694 & 0.541 & 0.635 & 0.539 & 0.639 & 0.550 & 0.713 \\ \cline{3-12}
         &  & ROC & 0.727 & 0.568 & 0.759 & 0.558 & 0.684 & 0.560 & 0.688 & 0.568 & 0.782 \\ \cline{3-12}
         &  & PR & 0.570 & 0.550 & 0.605 & 0.538 & 0.523 & 0.553 & 0.510 & 0.546 & 0.647 \\ \cline{2-12}
         &  \multirow{3}{0pt} {SSAST} & UAR & 0.681 & 0.555 & 0.696 & 0.551 & 0.652 & 0.546 & 0.662 & 0.555 & 0.725 \\ \cline{3-12}
         &  & ROC & 0.750 & 0.577 & 0.781 & 0.569 & 0.714 & 0.571 & 0.723 & 0.568 & 0.809 \\ \cline{3-12}
         &  & PR & 0.607 & 0.553 & 0.648 & 0.552 & 0.563 & 0.557 & 0.561 & 0.557 & 0.701 \\ \cline{2-12}
         & \multirow{3}{0pt}{BNN} & UAR & 0.678 & 0.558 & 0.696 & 0.551 & 0.657 & 0.558 & 0.660 & 0.535 & 0.716 \\ \cline{3-12}
         &  & ROC & 0.751 & 0.578 & 0.786 & 0.578 & 0.713 & 0.578 & 0.720 & 0.558 & 0.807 \\ \cline{3-12}
         & & PR & 0.601 & 0.550 & 0.647 & 0.556 & 0.551 & 0.554 & 0.563 & 0.551 & 0.691 \\ 
         \specialrule{.125em}{.1em}{.1em}
        \multirow{9}{0pt} {Cough} & \multirow{3}{0pt} {SVM} & UAR & 0.648 & 0.536 & 0.685 & 0.540 & 0.633 & 0.541 & 0.638 & 0.538 & 0.695 \\ \cline{3-12}
         &  & ROC & 0.712 & 0.544 & 0.748 & 0.550 & 0.687 & 0.559 & 0.692 & 0.559 & 0.763 \\ \cline{3-12}
         &  & PR & 0.559 & 0.526 & 0.594 & 0.535 & 0.533 & 0.550 & 0.521 & 0.545 & 0.625 \\ \cline{2-12}
         & \multirow{3}{0pt} {SSAST} & UAR & 0.681 & 0.545 & 0.690 & 0.541 & 0.638 & 0.528 & 0.640 & 0.543 & 0.702 \\ \cline{3-12}
         &  & ROC & 0.742 & 0.561 & 0.768 & 0.559 & 0.692 & 0.552 & 0.692 & 0.560 & 0.790 \\ \cline{3-12}
         &  & PR & 0.603 & 0.540 & 0.631 & 0.548 & 0.535 & 0.545 & 0.532 & 0.550 & 0.675 \\\cline{2-12}
         & \multirow{3}{0pt}{BNN} & UAR & 0.647 & 0.540 & 0.661 & 0.534 & 0.618 & 0.532 & 0.638 & 0.541 & 0.672 \\ \cline{3-12}
         &  & ROC & 0.732 & 0.570 & 0.765 & 0.563 & 0.682 & 0.542 & 0.698 & 0.556 & 0.786 \\ \cline{3-12}
         & & PR & 0.581 & 0.556 & 0.621 & 0.549 & 0.511 & 0.526 & 0.522 & 0.541 & 0.678 \\ 
          \specialrule{.125em}{.1em}{.1em}
        \multirow{9}{0pt} {Exhal ation} & \multirow{3}{0pt} {SVM} & UAR & 0.600 & 0.523 & 0.639 & 0.544 & 0.587 & 0.528 & 0.585 & 0.529 & 0.653 \\ \cline{3-12}
         &  & ROC & 0.646 & 0.555 & 0.690 & 0.559 & 0.618 & 0.541 & 0.621 & 0.550 & 0.712 \\ \cline{3-12}
         &  & PR & 0.477 & 0.560 & 0.513 & 0.547 & 0.444 & 0.536 & 0.431 & 0.543 & 0.566 \\ \cline{2-12}
         & \multirow{3}{0pt} {SSAST} & UAR & 0.649 & 0.553 & 0.663 & 0.558 & 0.593 & 0.531 & 0.588 & 0.531 & 0.660 \\ \cline{3-12}
         &  & ROC & 0.701 & 0.581 & 0.725 & 0.580 & 0.653 & 0.552 & 0.644 & 0.556 & 0.750 \\ \cline{3-12}
         &  & PR & 0.563 & 0.578 & 0.575 & 0.561 & 0.496 & 0.548 & 0.473 & 0.549 & 0.634 \\ \cline{2-12}
         & \multirow{3}{0pt}{BNN} & UAR & 0.576 & 0.529 & 0.581 & 0.526 & 0.603 & 0.525 & 0.601 & 0.541 & 0.608 \\ \cline{3-12}
         &  & ROC & 0.683 & 0.569 & 0.722 & 0.578 & 0.679 & 0.570 & 0.675 & 0.567 & 0.744 \\ \cline{3-12}
         & & PR & 0.539 & 0.581 & 0.573 & 0.563 & 0.519 & 0.573 & 0.507 & 0.551 & 0.620 \\  
\specialrule{.2em}{.1em}{.1em}
    \end{tabularx}
    \caption{Results detailing the performance of the \gls{svm}, \gls{ssast} and BNN models on the nine evaluation tasks for each of the four audio modalities: sentence, three coughs, cough and exhalation. The metrics which correspond to the highest performance across all modalities and models are \textbf{emboldened}. Each training and test set is shown with the corresponding support of \gls{covidpos} and COVID$^{-}$\xspace individuals. ROC and PR are used to abbreviate ROC-AUC and PR-AUC respectively.}
    \label{tab:main-results}
\end{table}

\clearpage

\section*{Figures}
\begin{figure}[h]
\captionsetup[subfigure]{labelformat=empty}
\centering
\centerline{
\begin{subfigure}{1\textwidth}
    \includegraphics[width=0.99\textwidth]{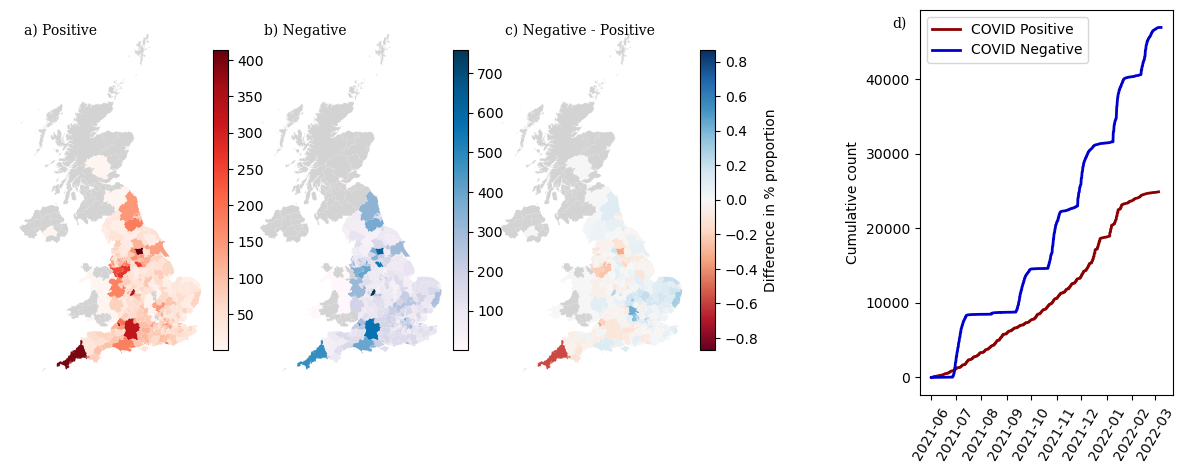}
\label{fig:demograph}
\end{subfigure}
}
\centerline{
\begin{subfigure}[b]{0.5\textwidth}
    \includegraphics[width=\textwidth]{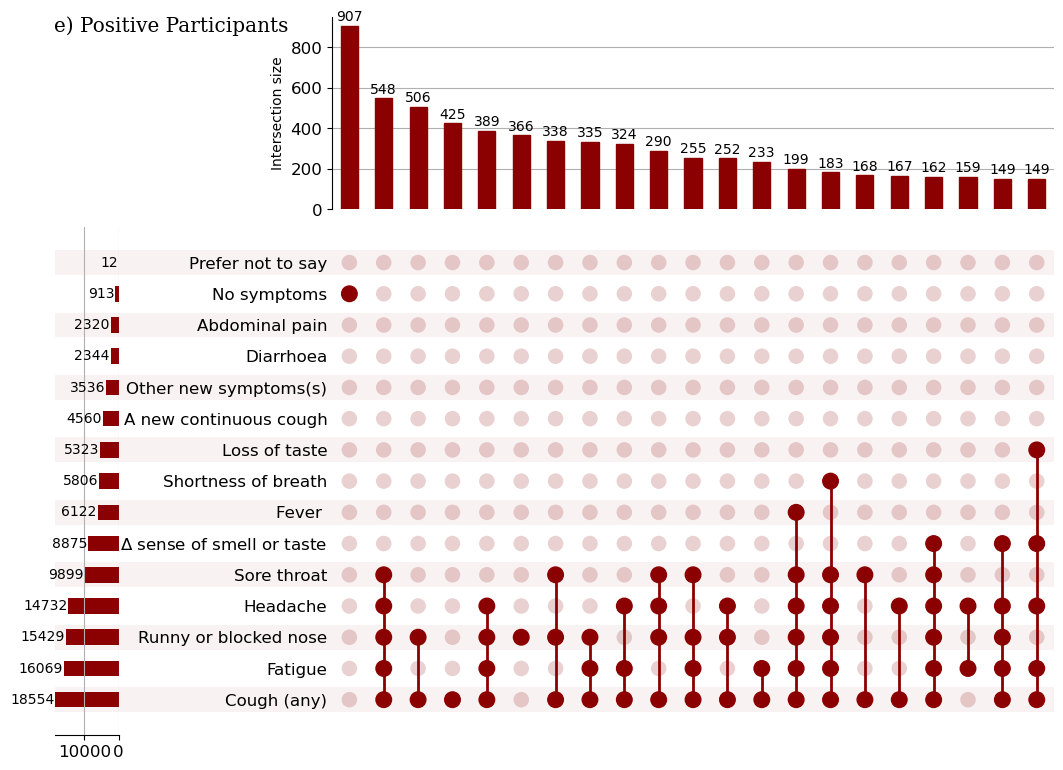}
    \label{fig:positiveupset}
\end{subfigure}
\begin{subfigure}[b]{0.5\textwidth}
    \includegraphics[width=\textwidth]{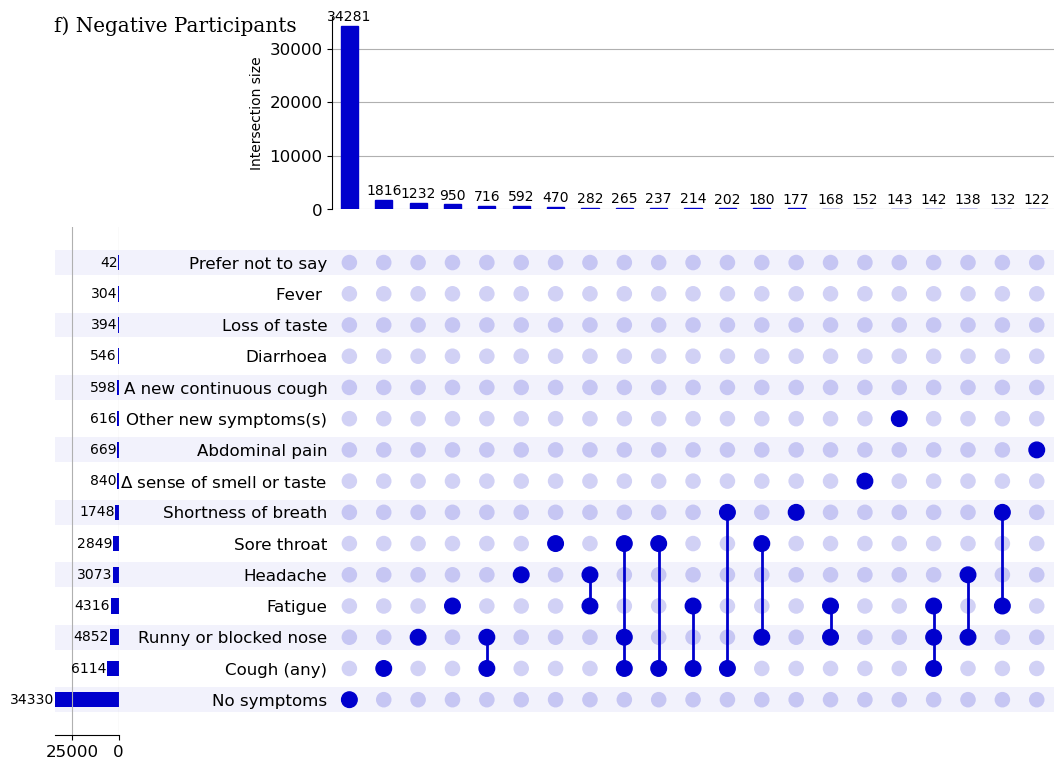}
    \label{fig:negativeupset}
\end{subfigure}
}
\centering
\centerline{
\begin{subfigure}{\textwidth}
    \subfloat[][{\fontfamily{ptm}\selectfont \scriptsize g) Data Filtration}]{\includegraphics[width=0.5\textwidth]{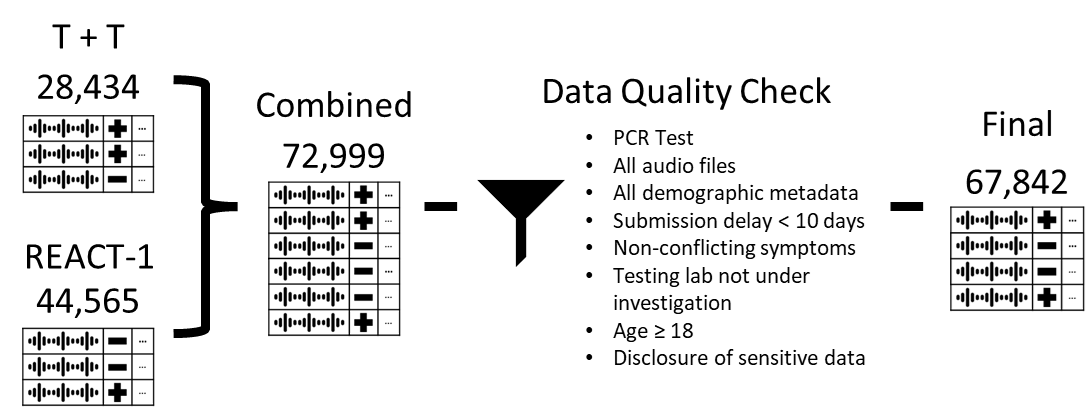}} \hspace{0.5cm}
    \subfloat[][{\fontfamily{ptm}\selectfont \scriptsize h) Data Splits}]{\resizebox{0.5\textwidth}{!}{
\begin{tabular}{cc||c|c|c|c|c}
\specialrule{.2em}{.1em}{.1em}
    ~ & ~ & Random & Standard & Matched & Long & Long Matched \\ \hline
    ~ & $+$ & 20000 & 9379 & 2599 & ~ & ~ \\ 
    train & $-$ & 37665 & 16518 & 2599 & ~ & ~ \\ 
    ~ & $\sum$ & 57665 & 25897 & 5198 & ~ & ~ \\ \hline
    ~ & $+$ & 3514 & 3820 & 907 & 10315 & 2098 \\ 
    test & $-$ & 6663 & 7301 & 907 & 20509 & 2098 \\ 
    ~ & $\sum$ & 10177 & 11121 & 1814 & 30824 & 4196 \\
\specialrule{.2em}{.1em}{.1em}
\end{tabular}

\label{tab:splits}
}}
\end{subfigure}
}
\caption{Demographic statistics of collected dataset a-b) geographical locations of COVID positive and negative \gls{pcr} confirmed participants. c) $\% \frac{100 \times \text{\# negative participants at location}}{\text{\# negative participants in total}} - \frac{100 \times \text{\# positive participants at location}}{\text{\# positive participants in total}}$
 d) Cumulative count of the \# participants partaking in the study. e-f) the 21 most common combinations of symptoms for COVID positive, e) and negative, f) participants. Ordered along the x-axis by total number of participants displaying that particular combination of symptoms. Symptoms are ordered along the y-axis according to total number of participants displaying at least that symptom at the time of recording. g) Schematic detailing the two recruitment sources for the study and the filtration steps applied to yield the final dataset. h) Dataset splits in participant numbers}
\label{fig:demographics}
\end{figure}


\begin{figure}[t]
	\vspace{.4cm}
\begin{center}
\small
\newcommand\predcovsymlab{\begin{tabular}{c}Predict COVID from\\binary symptoms\end{tabular}}
\resizebox{1\textwidth}{!}{

    \begin{tikzpicture}
        \def\minsize{1.25cm}
        \def\mintextwidth{8cm}
		\def\xlab{2.5cm}
		\def\ylab{1.2cm}
		\def\yshift{0.25cm}
		\def\yshiftmvuv{0.25cm}
		\def\xshift{4cm}
        \newcommand{\toplabx}{0cm}
        \newcommand{\toplaby}{-.75cm}
        \newcommand{\bottomlaby}{.75cm}
 		\node[const] (tab2){
\begin{tabular}{rrr}
  \hline
 & COVID+ & COVID- \\ 
  \hline
Symptomatic & 33.4\% & 17.5\% \\ 
  Asymptomatic & 1.3\% & 47.8\% \\ 
   \hline
\end{tabular}
};
         \node[const, above=of tab2, xshift=\toplabx,yshift = \toplaby] (btitle) {(a)  Symptoms-based enrolment };
        \node[const, below=of tab2, yshift = \bottomlaby] (bsubtitle) {$
                  \predcovsymlab
                 \left\{\begin{tabular}{r}
                      \footnotesize{$\rho$=\input{text_numbers/r2_us.txt}}\\
                        \footnotesize{MI=\input{text_numbers/mi_us.txt}}\\
                        \footnotesize{Sensitivity=\input{text_numbers/sens_sym_us.txt}}\\
              \footnotesize{Specificity=\input{text_numbers/spec_sym_us.txt}}\\
                            \footnotesize{AUC=\input{text_numbers/auc_us.txt}}\unskip
                        \end{tabular}\right.$};
       	\node[const, right=of tab2, xshift = 0cm] (tab1){
\begin{tabular}{rrr}
  \hline
 & COVID+ & COVID- \\ 
  \hline
Symptomatic & 1.3\% & 19.6\% \\ 
  Asymptomatic & 0.7\% & 78.4\% \\ 
   \hline
\end{tabular}
};
        \node[const, above=of tab1, xshift=\toplabx,yshift = \toplaby] (atitle) {(b)  General population enrolment };
        \node[const, below=of tab1, yshift = \bottomlaby] (asubtitle) {$
                  \predcovsymlab
                 \left\{\begin{tabular}{r}
                        \footnotesize{$\rho$=\input{text_numbers/r2_genpop.txt}}\\
                        \footnotesize{MI=\input{text_numbers/mi_genpop.txt}}\\
                      \footnotesize{Sensitivity=\input{text_numbers/sens_sym_genpop.txt}}\\
                      \footnotesize{Specificity=\input{text_numbers/spec_sym_genpop.txt}}\\
                        \footnotesize{AUC=\input{text_numbers/auc_genpop.txt}}\unskip
                    \end{tabular}\right.$};
 		\node[const, right=of tab1, xshift = 0cm] (tab3){
\begin{tabular}{rrr}
  \hline
 & COVID+ & COVID- \\ 
  \hline
Symptomatic & 46.5\% & 46.5\% \\ 
  Asymptomatic & 3.5\% & 3.5\% \\ 
   \hline
\end{tabular}
};
         \node[const, above=of tab3, xshift=\toplabx,yshift = \toplaby] (btitle) {(c)  Matched enrolment };
        \node[const, below=of tab3, yshift = \bottomlaby] (bsubtitle) {$
                  \predcovsymlab
                 \left\{\begin{tabular}{r}
                          \footnotesize{$\rho$=\input{text_numbers/r2_matched.txt}}\\
                        \footnotesize{MI=\input{text_numbers/mi_matched.txt}}\\
                        \footnotesize{Sensitivity=\input{text_numbers/sens_sym_matched.txt}}\\
              \footnotesize{Specificity=\input{text_numbers/spec_sym_matched.txt}}\\
                            \footnotesize{AUC=\input{text_numbers/auc_matched.txt}}\unskip
                    \end{tabular}\right.$};

	\end{tikzpicture}
}
\end{center}
\caption{Illustrative 2-by-2 tables relating symptoms status with \gls{covid} status. (a) Symptoms-based enrolment in which \gls{covidpos} individuals are preferentially recruited based upon exhibiting symptoms (percentages are calculated from the entire sample of individuals recruited into this study). (b) General population enrolment, based on random sampling from an illustrative general population with \gls{covid} prevalence 2\%, and with symptomatic individuals making up 20\% and 65\% of \gls{covidneg} and  \gls{covidpos} sub-populations respectively. (c) Matched enrolment, in which the number of \gls{covidneg} and \gls{covidpos} individuals is the same for each particular symptoms profile, here within the symptomatic and asymptomatic subgroups (percentages shown are for the Matched test set in the current study). For each type of enrolment, diagnostic accuracy of the resulting symptoms-only \gls{covid} classifier are shown below the table: $\rho$ (population correlation coefficient), \gls{mi}, sensitivity, specificity and AUC.}
	\label{fig:bias_intro}
\end{figure}

        \def\xbetween{14.5cm}
        \def\xbetweentwice{31cm}
        \def\ybetween{8.85cm}
        \def\thick{2mm}
        \def\ymidshift{2.125cm}
        \def\xmidshift{1.25cm}
        \newcommand\incother{0}
        \newcommand\incplate{1}
        \newcommand\covidsignaturelinestyle{dashed}
        \newcommand\measconf{\begin{tabular}{c}
		        Self-reported  \\
		        symptoms\\
		        $\zlab$
		    \end{tabular}}
		    
	    \newcommand\otherconf{\begin{tabular}{c}
		      Other\\
		      confounders,\\
		      age,location\\
		      $\vlab$
		    \end{tabular}}
	    \newcommand\agesex{\begin{tabular}{c}
		      Age, sex
		    \end{tabular}}
		    
	    \newcommand\covidsignature{\begin{tabular}{c}
		        COVID\\
		        acoustic\\
		        signature\\
		        $\wlab$
		    \end{tabular}}
	    \newcommand\acousticdata{\begin{tabular}{c}
		      Acoustic\\
		      recording\\
		      $\xlab$
		    \end{tabular}}
	    \newcommand\acousticclassifier{\begin{tabular}{c}
		      Trained\\
		      acoustic\\
		      classifier\\
		      $\predlab$
		    \end{tabular}}
	    \newcommand\covidstatus{\begin{tabular}{c}
		        \gls{covid}  \\
		        status\\
		        $\ylab$
		    \end{tabular}}
    \newcommand\platedag{
	    \if\incplate1
    		\plate[inner sep=.3cm] {plate1}{(x)(z)(y)(s)(e)} {\LARGE$i$};
         \else \fi
         }

    \newcommand\platecigraph{
	    \if\incplate1
    		\plate[inner sep=.3cm] {plate1}{(x)(z)(y)(s)(e)} {\LARGE$i: \elab=1$};
         \else \fi
         }

\begin{figure}[t]
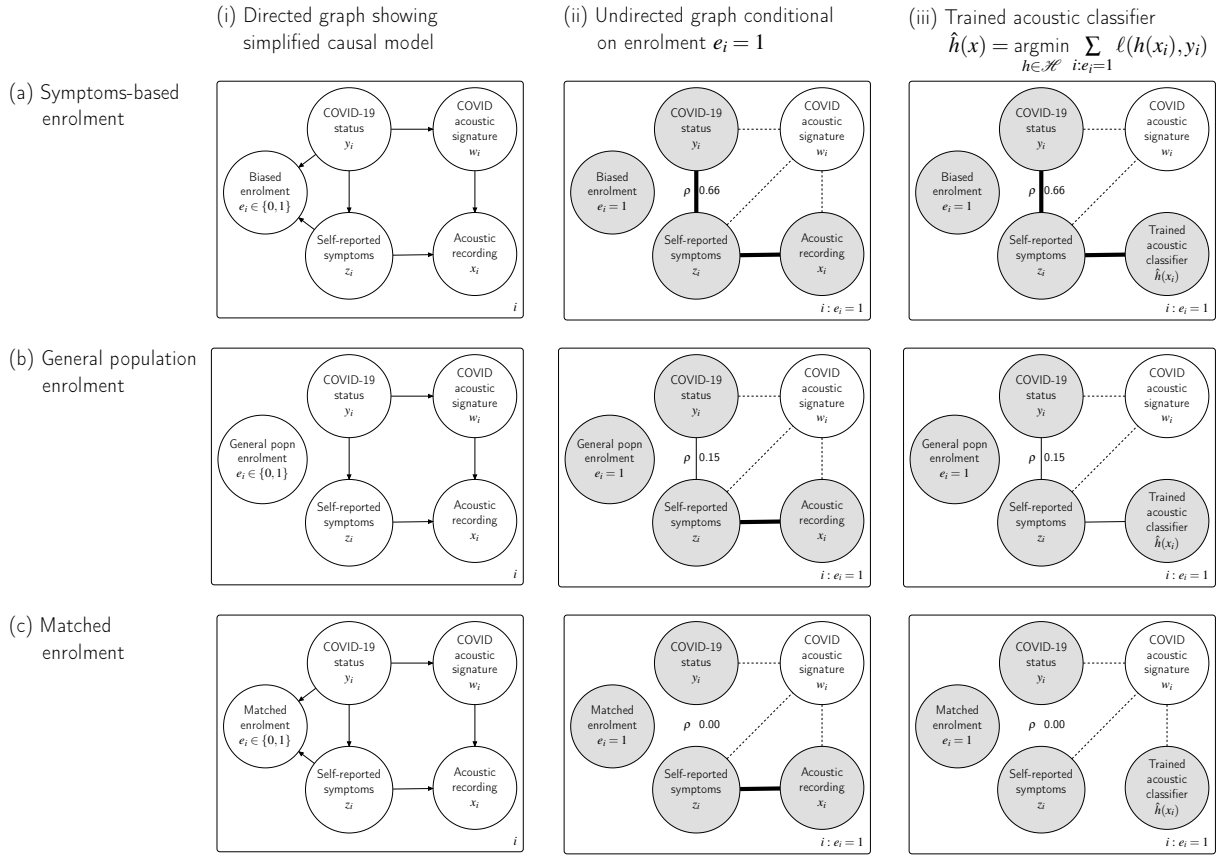

	\vspace{.4cm}
\begin{center}
\usetikzlibrary{positioning, matrix,shapes,arrows,fit,automata}
\def\minsize{4cm}
 \tikzstyle{latent} = [circle,fill=white,draw=black,inner sep=1pt,
            minimum size=20pt,minimum size=\minsize, font=\fontsize{17}{17}\selectfont, node distance=2]

\resizebox{1\textwidth}{!}{
        \begin{tikzpicture}[node distance=7cm and 2cm,auto]
        \def\vthickline{5pt}
        \def\thickline{2.5pt}
        \def\normalline{1pt}
        \node[latent] (y) {\covidstatus};
        \node[const, right=0cm of y] (y1) {};
        \node[latent, below=of y] (z) {\measconf};
        \node[latent, left=of $(y)!0.5!(z)$] (e) {\begin{tabular}{c}
		        Biased\\
		        enrolment\\
		        $\elab\in \{0,1\}$
		    \end{tabular}};
        \node[latent, right=of y, yshift = 0cm] (s) {\covidsignature};
        \node[latent, below=of s] (x) {\acousticdata};
    \if\incother1
       \node[latent, right=\xmidshift of $(y)!0.5!(z)$] (v) {\otherconf};
    		\edge {v} {y}; 
	    	\edge {v} {x}; 
      \else \fi
	
		\edge {z} {x}; 
		\edge {y} {z}; 
		\edge {y} {s}; 
		\edge {y} {e}; 
		\edge {z} {e}; 
		\edge {s} {x}; 
		\platedag

        \node[latent,below=0cm of y, yshift = -\ybetween] (y) {\covidstatus};
        \node[const, right=0cm of y] (y2) {};
        \node[latent, below=of y] (z) {\measconf};
        \node[latent, right=of y, yshift = 0cm] (s) {\covidsignature};
        \node[latent, below=of s] (x) {\acousticdata};
       \node[latent, left=of $(y)!0.5!(z)$] (e) {\begin{tabular}{c}
		        General popn\\
		        enrolment\\
		        $\elab\in \{0,1\}$
		    \end{tabular}};
        \if\incother1
           \node[latent, right=\xmidshift of $(y)!0.5!(z)$] (v) {\otherconf};
    		\edge {v} {y}; 
	    	\edge {v} {x}; 
      \else
         \fi
		\edge {z} {x}; 
		\edge {y} {z}; 
		\edge {y} {s}; 
		\edge {s} {x}; 
        \platedag

        \node[latent,below=0cm of y, yshift = -\ybetween] (y) {\covidstatus};
        \node[const, right=0cm of y] (y3) {};
        \node[latent, below=of y] (z) {\measconf};
        \node[latent, left=of $(y)!0.5!(z)$] (e) {\begin{tabular}{c}
		        Matched\\
		        enrolment\\
		        $\elab\in \{0,1\}$
		    \end{tabular}};
		    
        \node[latent, right=of y, yshift = 0cm] (s) {\covidsignature};
        \node[latent, below=of s] (x) {\acousticdata};
    \if\incother1
           \node[latent, right=\xmidshift of $(y)!0.5!(z)$] (v) {\otherconf};
    		\edge {v} {y}; 
	    	\edge {v} {x}; 
      \else
         \fi
		\edge {z} {x}; 
		\edge {y} {z}; 
		\edge {y} {s}; 
		\edge {y} {e}; 
		\edge {z} {e}; 
		\edge {s} {x}; 
        \platedag

    	\newcommand\rlababovez{1cm}	
    	\newcommand\rlabxshift{-.6cm}	
    	\newcommand\rlabanchor{west}	
    	\newcommand\labeledgeus{\node[const, above=\rlababovez of z, rotate=0,xshift =\rlabxshift,anchor=\rlabanchor] {\scalebox{1.6}{$\rho$\ \ \input{text_numbers/r2_us.txt}}}}
    	\newcommand\labeledgegenpop{\node[const, above=\rlababovez of z, rotate=0,xshift =\rlabxshift,anchor=\rlabanchor] {\scalebox{1.6}{$\rho$\ \ \input{text_numbers/r2_genpop.txt}}}}
    	\newcommand\labeledgematched{\node[const, above=\rlababovez of z, rotate=0,xshift =\rlabxshift,anchor=\rlabanchor] {\scalebox{1.6}{$\rho$\ \ \input{text_numbers/r2_matched.txt}}}}
    	
        \node[obs,right of= y1, xshift = \xbetween] (y) {\covidstatus};
        \node[const, right=0cm of y] (y12) {};
        \node[obs, below=of y] (z) {\measconf};
        \node[latent, right=of y, yshift = 0cm] (s) {\covidsignature};
        \node[obs, below=of s] (x) {\acousticdata};
        \node[obs, left=of $(y)!0.5!(z)$] (e) {\begin{tabular}{c}
		        Biased\\
		        enrolment\\$\elab=1$
		    \end{tabular}};
        \if\incother1
            \node[latent, right=\xmidshift of $(y)!0.5!(z)$] (v) {\otherconf};
            \edge[-] {y} {v}; 
            \edge[-] {v} {x}; 
        \else
        \fi
		\edge[-,line width=\thick] {z} {x}; 
		\edge[-,line width=\thick] {y} {z}; 
		\edge[-,\covidsignaturelinestyle] {z} {s}; 
		\edge[-,\covidsignaturelinestyle] {y} {s}; 
		\edge[-,\covidsignaturelinestyle] {s} {x}; 
		\node[const, above=\rlababovez of z, rotate=0,xshift =\rlabxshift,anchor=\rlabanchor] {\scalebox{1.6}{$\rho$\ \ \input{text_numbers/r2_us.txt}}};
        \platecigraph
        
        \node[obs,right of= y2, xshift = \xbetween] (y) {\covidstatus};
        \node[obs, below=of y] (z) {\measconf};
        \node[latent, right=of y, yshift = 0cm] (s) {\covidsignature};
        \node[obs, below=of s] (x) {\acousticdata};
        \node[obs, left=of $(y)!0.5!(z)$] (e) {\begin{tabular}{c}
		        General popn\\
		        enrolment\\$\elab=1$
		    \end{tabular}};
        \if\incother1
            \node[latent, right=\xmidshift of $(y)!0.5!(z)$] (v) {\otherconf};
            \edge[-] {y} {v}; 
            \edge[-] {v} {x}; 
        \else
        \fi
 		\edge[-,line width=\thick] {z} {x}; 
		\edge[-] {y} {z}; 
		\edge[-,\covidsignaturelinestyle] {z} {s}; 
		\edge[-,\covidsignaturelinestyle] {y} {s}; 
		\edge[-,\covidsignaturelinestyle] {s} {x}; 
	    \node[const, above=\rlababovez of z, rotate=0,xshift =\rlabxshift,anchor=\rlabanchor] {\scalebox{1.6}{$\rho$\ \ \input{text_numbers/r2_genpop.txt}}};
        \platecigraph

        \node[obs,right of= y3, xshift = \xbetween] (y) {\covidstatus};
        \node[obs, below=of y] (z) {\measconf};
        \node[latent, right=of y, yshift = 0cm] (s) {\covidsignature};
        \node[obs, below=of s] (x) {\acousticdata};
        \node[obs, left=of $(y)!0.5!(z)$] (e) {\begin{tabular}{c}
		        Matched\\
		        enrolment\\$\elab=1$
		    \end{tabular}};
        \if\incother1
            \node[latent, right=\xmidshift of $(y)!0.5!(z)$] (v) {\otherconf};
            \edge[-] {y} {v}; 
            \edge[-] {v} {x}; 
        \else
        \fi
		\edge[-,line width=\thick] {z} {x}; 
		\edge[-,\covidsignaturelinestyle] {z} {s}; 
		\edge[-,\covidsignaturelinestyle] {y} {s}; 
		\edge[-,\covidsignaturelinestyle] {s} {x}; 
		\node[const, above=\rlababovez of z, rotate=0,xshift =\rlabxshift,anchor=\rlabanchor] {\scalebox{1.6}{$\rho$\ \ \input{text_numbers/r2_matched.txt}}};
        \platecigraph

        \node[obs,right of= y1, xshift = \xbetweentwice] (y) {\covidstatus};
        \node[const, right=0cm of y] (y13) {};
        \node[obs, below=of y] (z) {\measconf};
        \node[latent, right=of y, yshift = 0cm] (s) {\covidsignature};
        \node[obs, below=of s] (x) {\acousticclassifier};
        \node[obs, left=of $(y)!0.5!(z)$] (e) {\begin{tabular}{c}
		        Biased\\
		        enrolment\\$\elab=1$
		    \end{tabular}};
        \if\incother1
            \node[latent, right=\xmidshift of $(y)!0.5!(z)$] (v) {\otherconf};
            \edge[-] {y} {v}; 
            \edge[-] {v} {x}; 
        \else
        \fi
 		\edge[-,line width=\thick] {z} {x}; 
		\edge[-,line width=\thick] {y} {z}; 
		\edge[-,\covidsignaturelinestyle] {z} {s}; 
		\edge[-,\covidsignaturelinestyle] {y} {s}; 
		\node[const, above=\rlababovez of z, rotate=0,xshift =\rlabxshift,anchor=\rlabanchor] {\scalebox{1.6}{$\rho$\ \ \input{text_numbers/r2_us.txt}}};
        \platecigraph
        
        \node[obs,right of= y2, xshift = \xbetweentwice] (y) {\covidstatus};
        \node[obs, below=of y] (z) {\measconf};
        \node[latent, right=of y, yshift = 0cm] (s) {\covidsignature};
        \node[obs, below=of s] (x) {\acousticclassifier};
        \node[obs, left=of $(y)!0.5!(z)$] (e) {\begin{tabular}{c}
		        General popn\\
		        enrolment\\$\elab=1$
		    \end{tabular}};
        \if\incother1
            \node[latent, right=\xmidshift of $(y)!0.5!(z)$] (v) {\otherconf};
            \edge[-] {y} {v}; 
            \edge[-] {v} {x}; 
        \else
        \fi
 		\edge[-] {z} {x}; 
		\edge[-] {y} {z}; 
		\edge[-,\covidsignaturelinestyle] {z} {s}; 
		\edge[-,\covidsignaturelinestyle] {y} {s}; 
		\node[const, above=\rlababovez of z, rotate=0,xshift =\rlabxshift,anchor=\rlabanchor] {\scalebox{1.6}{$\rho$\ \ \input{text_numbers/r2_genpop.txt}}};
        \platecigraph
        
        \node[obs,right of= y3, xshift = \xbetweentwice] (y) {\covidstatus};
        \node[obs, below=of y] (z) {\measconf};
        \node[latent, right=of y, yshift = 0cm] (s) {\covidsignature};
        \node[obs, below=of s] (x) {\acousticclassifier};
        \node[obs, left=of $(y)!0.5!(z)$] (e) {\begin{tabular}{c}
		        Matched\\
		        enrolment\\$\elab=1$
		    \end{tabular}};
        \if\incother1
            \node[latent, right=\xmidshift of $(y)!0.5!(z)$] (v) {\otherconf};
            \edge[-] {y} {v}; 
            \edge[-] {v} {x}; 
        \else
        \fi
		\edge[-,\covidsignaturelinestyle] {y} {s}; 
 		\edge[-,\covidsignaturelinestyle] {s} {x}; 
		\edge[-,\covidsignaturelinestyle] {z} {s}; 
		\node[const, above=\rlababovez of z, rotate=0,xshift =\rlabxshift,anchor=\rlabanchor] {\scalebox{1.6}{$\rho$\ \ \input{text_numbers/r2_matched.txt}}};
        \platecigraph

        \newcommand\tabtitle{1.3cm}
        \newcommand\tabtitlerow{1.6cm}
        \newcommand\fontsizetitle{30}
        \newcommand\rowlabyshift{1.1cm}
        \newcommand\rowlabxshift{-17.5cm}
        \newcommand\collabyshift{3.75cm}
        \newcommand\collabxshift{-8.6cm}
        \node[const,above=of y1,font=\fontsize{\fontsizetitle}{\fontsizetitle}\selectfont,yshift = \collabyshift,xshift = \collabxshift,anchor=west] (title1) {\begin{tabular}{l}
		      (i) Directed graph showing\\
		      \hspace{\tabtitle}simplified causal model
	    \end{tabular}};
		\node[const,above=of y12,font=\fontsize{\fontsizetitle}{\fontsizetitle}\selectfont,yshift = \collabyshift,xshift = \collabxshift,anchor=west] (title2) {\begin{tabular}{l}
		      (ii) Undirected graph conditional\\
		      \hspace{\tabtitle}\ on enrolment $\elab=1$
	    \end{tabular}};
		\node[const,above=of y13,font=\fontsize{\fontsizetitle}{\fontsizetitle}\selectfont,yshift = \collabyshift-.5cm,xshift = \collabxshift,anchor=west] (title3) {\begin{tabular}{l}
		      (iii) Trained acoustic classifier\\
		\hspace{\tabtitle}$\ \ \hat{h}(x)= \underset{h\in \mathcal{H}}{\text{argmin}}\sum\limits_{i:\elab=1} \ell (h(x_i), y_i)$
	    \end{tabular}};
		\node[const,left=of y1,font=\fontsize{\fontsizetitle}{\fontsizetitle}\selectfont,yshift = \rowlabyshift,xshift = \rowlabxshift, rotate = 0,anchor=west] (title1) {\begin{tabular}{l}
		      (a) Symptoms-based\\
		      \hspace{\tabtitlerow}enrolment
	    \end{tabular}};
		\node[const,left=of y2,font=\fontsize{\fontsizetitle}{\fontsizetitle}\selectfont,yshift = \rowlabyshift,xshift = \rowlabxshift, rotate = 0,anchor=west] (title1) {\begin{tabular}{l}
		      (b) General population\\
		      \hspace{\tabtitlerow}enrolment
	    \end{tabular}};
		\node[const,left=of y3,font=\fontsize{\fontsizetitle}{\fontsizetitle}\selectfont,yshift = \rowlabyshift,xshift = \rowlabxshift, rotate = 0,anchor=west] (title1) {\begin{tabular}{l}
		      (c) Matched\\
		      \hspace{\tabtitlerow}enrolment
	    \end{tabular}};
	\end{tikzpicture}
}
\end{center}
\vspace{-.25cm}
    	\caption{Graphical representation of enrolment effects. Rows (a-c) presenting various types of enrolment and columns (i-iii) showing different types of graph, in which shaded nodes are observed variables. (a) Symptoms-based enrolment enforces a supervised sampling regime in which \gls{covidpos} individuals are preferentially recruited based upon exhibiting symptoms (e.g.\ Figure~\ref{fig:bias_intro}(a)). (b) Randomised enrolment performs random sampling of individuals from the general population.  (c) Matched enrolment balances the number of \gls{covidpos} and \gls{covidpos} individuals that share each particular symptoms profile. (i) Bayesian knowledge graphs displaying a simplified causal model. (ii) Undirected conditional independence graphs implied by the directed graphs in (i) when we condition on enrolment ($\elab=1$). (iii) Undirected conditional independence graphs, as in (ii), but now showing the trained acoustic classifier $\predlab$ (trained to predict $\ylab$ based on input $\xlab$) in lieu of the acoustic recording data $\xlab$. } 
\label{fig:causal_model}
\end{figure}

\begin{figure}[h!]
    \centering
    \includegraphics[width=\textwidth]{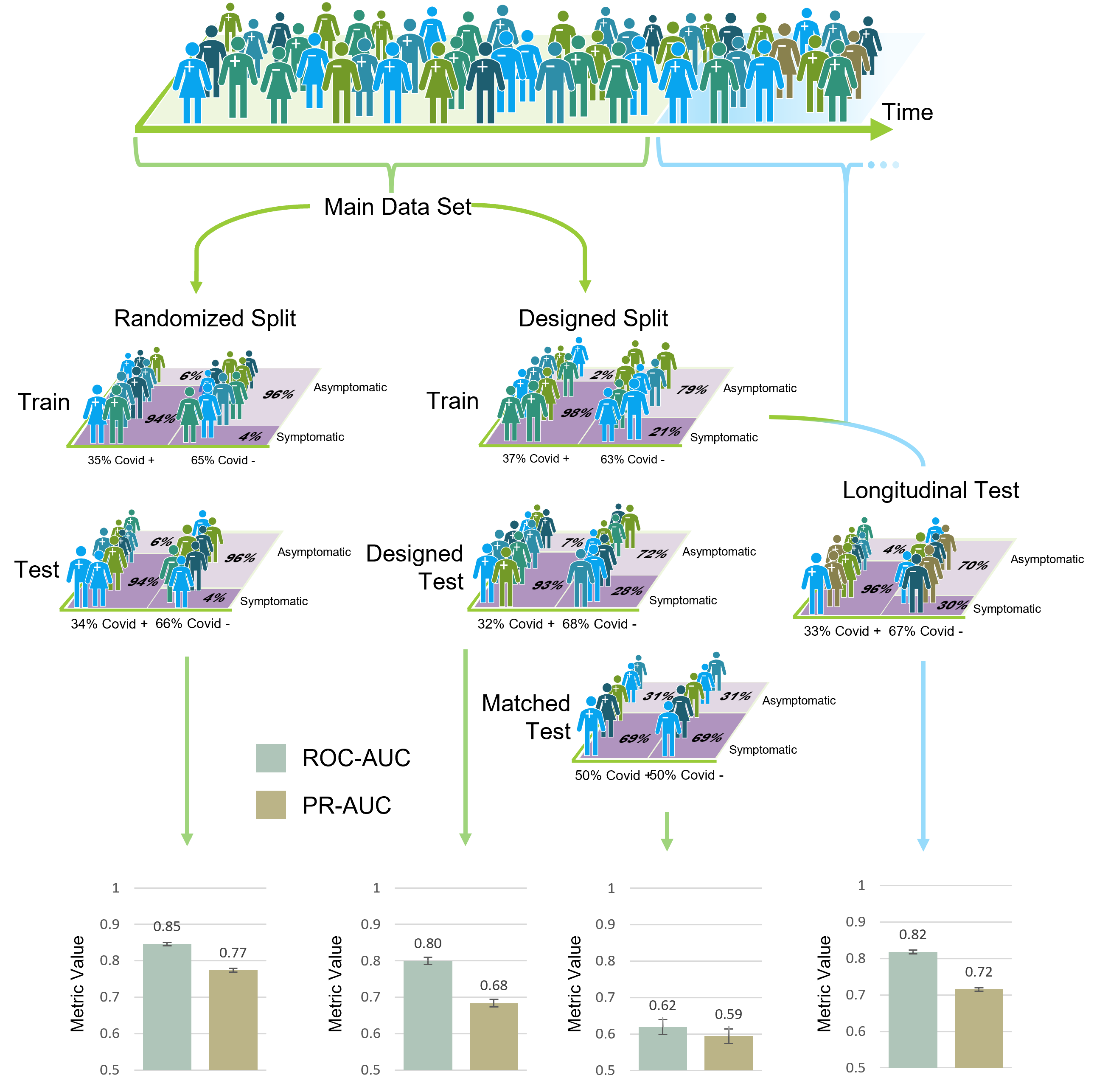}    \caption{Schematic demonstrating the importance of ascertainment bias adjustment in the context of reporting \gls{covid} detection from patient respiratory audio performance. Human figures represent study participants and their corresponding \gls{covid} infection status, with the different colours portraying different demographic or symptomatic features. When participants are randomly split into training and test sets, the ``Randomised Split'', models perform well at \gls{covid} detection, achieving AUCs in excess of 0.8. However, ``Matched Test'' set performance is seen to drop to estimated AUC between 0.60-0.65, with an AUC of 0.5 representing random classification. Inflated classification performance is also seen in engineered out of distribution test sets such as: the ``Designed Test'' set, where a select set of demographic groups appear solely in the test set, and the ``Longitudinal Test'' set where there is no overlap in the time of submission between train and test instances.}
    \label{fig:illustrative}
\end{figure}

\begin{figure}[h!]
    \centering
    \includegraphics[width=\textwidth]{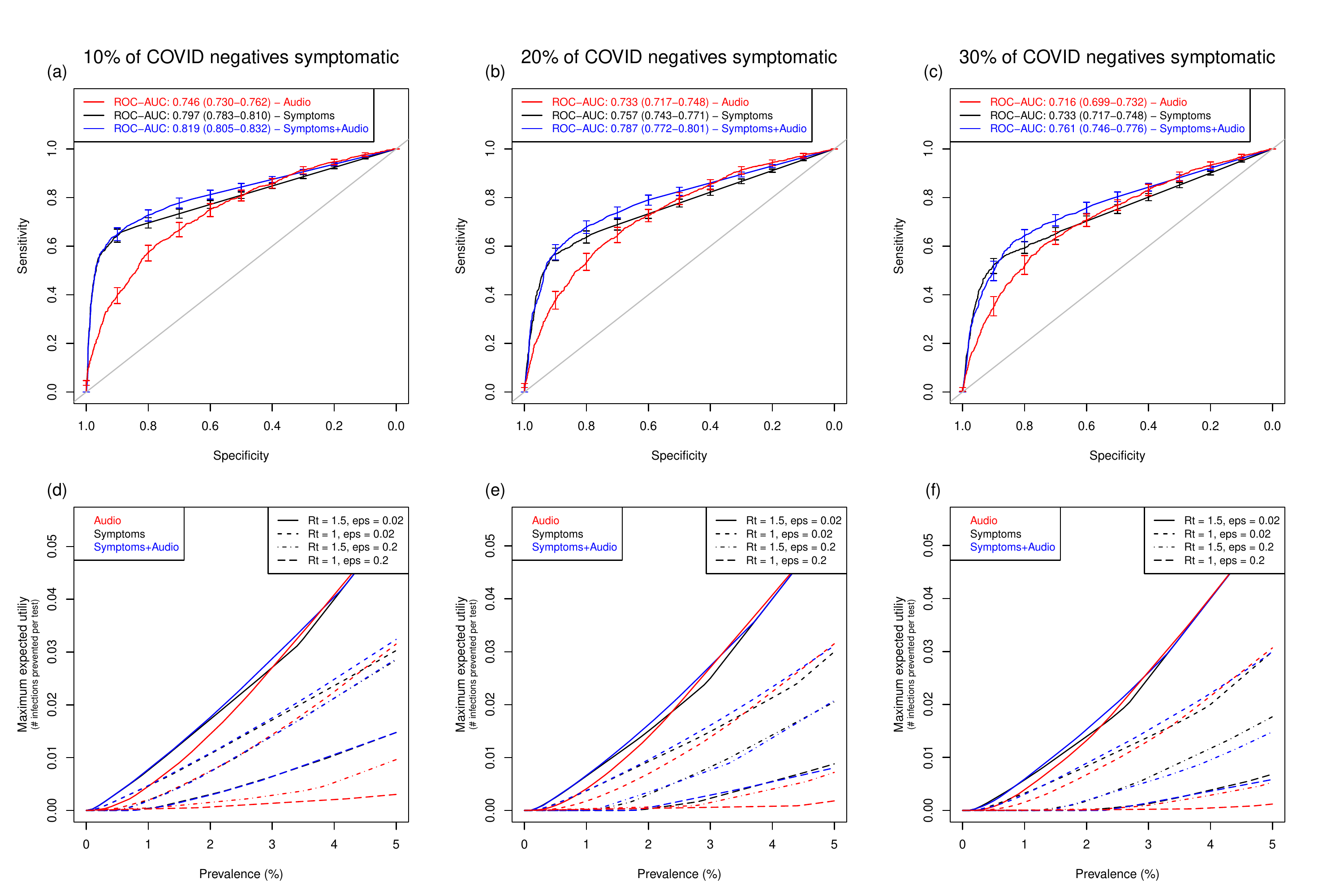}
    \caption{Comparison of sensitivity, specificity, and utility across audio-based and symptoms-based classifiers, as applied in a simulated general populations. The percentage of \gls{covidneg} individuals who are symptomatic in the general population varies between 10\% and 30\% across the three columns of panels (labelled top). (a)-(c) Comparison of ROC curves between the Audio, Symptoms, and Symptoms+Audio classifiers;  panel legends show curve colour for each classifier, along with \gls{roc} estimates and 95\% DeLong CIs.
    (d)-(f) Comparison of maximum expected utility across classifiers. Four different utility functions are included, as detailed in the top-right legend (utility function parameters $R_t$, $\varepsilon$ and $\delta$ are defined in Results; in this Figure, $\delta=0$). Curves are coloured to indicate Audio, Symptoms or Symptoms+Audio classifiers, as detailed in the top left legend.}
    \label{fig:audio_vs_symptoms_genpop}
\end{figure}

\clearpage

\bibliography{main}

\section*{Acknowledgements}
Authors gratefully acknowledge the contributions of staff from \gls{nhs} Test and Trace Lighthouse Labs, REACT Study, Ipsos MORI, Studio24, Fujitsu Services Ltd. Authors in The Alan Turing Institute and Royal Statistical Society Health Data Lab gratefully acknowledge funding from Data, Analytics and Surveillance Group, a part of the \gls{ukhsa}. This work was funded by The Department for Health and Social Care (Grant ref: 2020/045) with support from The Alan Turing Institute (EP/W037211/1) and in-kind support from The Royal Statistical Society. J.B. acknowledges funding from the i-sense EPSRC IRC in Agile Early Warning Sensing Systems for Infectious Diseases and Antimicrobial Resistance EP/R00529X/1. A.T.C. acknowledges funding from the European Union's Horizon 2020 research and innovation programme under the Marie Skłodowska–Curie grant agreement No 801604. SP acknowledges funding from the Economic and Social Research Council (ESRC) [grant number ES/P000592/1]. 

\clearpage

\section*{Supplementary Information}
\label{sec:SI}
\renewcommand{\thefigure}{S\arabic{figure}}
\setcounter{figure}{0}
\renewcommand{\thetable}{S\arabic{table}}
\setcounter{table}{0}

\subsection*{Supplementary Figures and Tables}

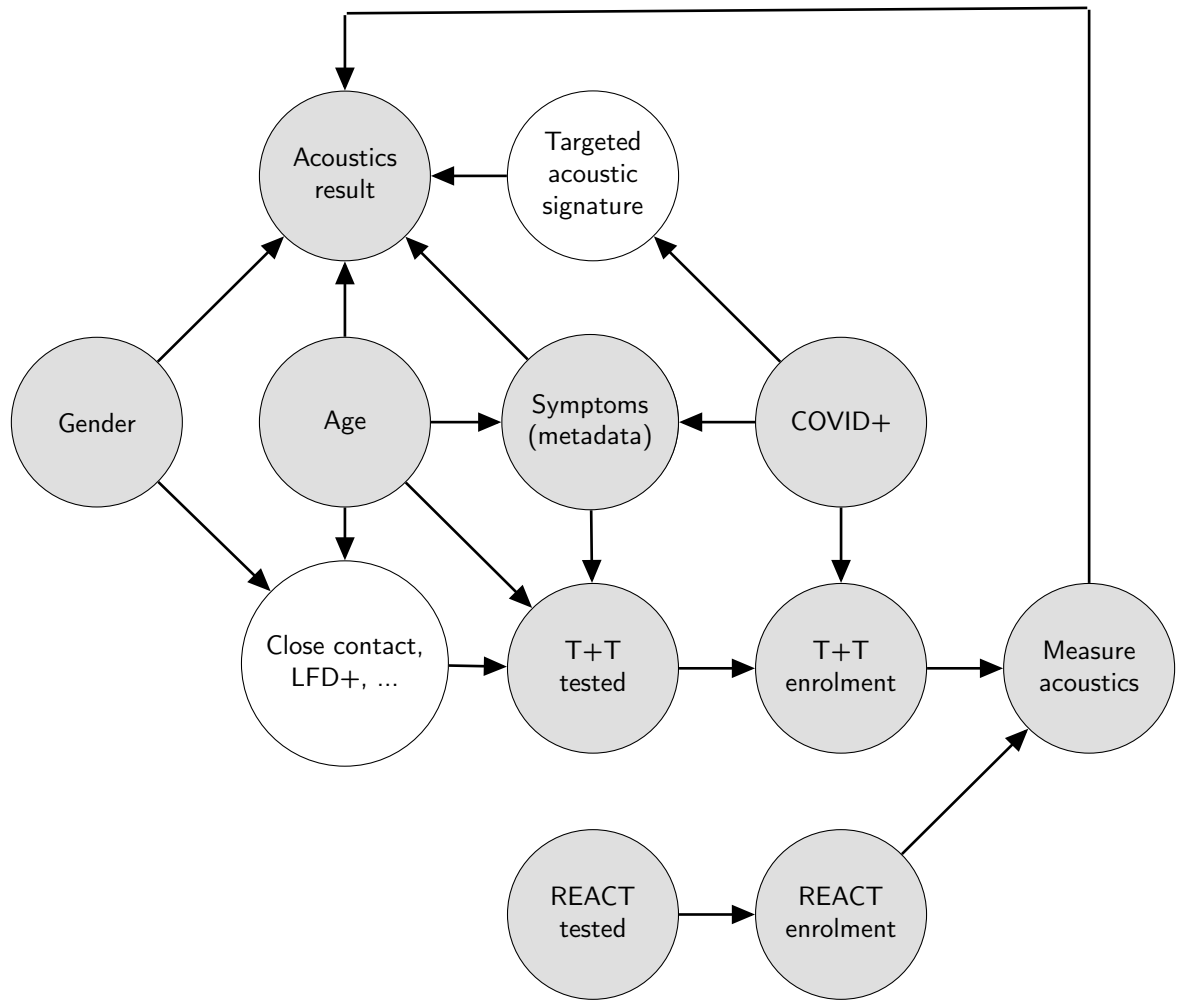
\begin{figure}[ht]
	\vspace{.4cm}
\begin{center}
        \def\minsize{2.25cm}
    \begin{tikzpicture}[mynode/.style={circle,draw=black,minimum size=\minsize}]
        \def\vthickline{5pt}
        \def\thickline{2.5pt}
        \def\normalline{1pt}
		\node[obs,minimum size=\minsize,xshift=-2cm] (covid) {COVID+};
		\node[obs,minimum size=\minsize, left=of covid] (sym) {Symptoms};
		\node[const,minimum size=\minsize, above=of covid] (ses) {};
		\node[obs,minimum size=\minsize, left=of sym] (age) {Age};
		\node[obs,minimum size=\minsize, left=of age] (sex) {Gender};
		\node[obs,minimum size=\minsize, above=of age] (acoustics) {\begin{tabular}{c}
		        Acoustics  \\
		        result
		    \end{tabular}};
		\node[obs,minimum size=\minsize, below=of sym] (tested) {\begin{tabular}{c}
		        \gls{tt}  \\
		        tested
		    \end{tabular}};
		\node[obs,minimum size=\minsize, below=of tested] (react) {\begin{tabular}{c}
		        \gls{react}  \\
		        tested
		    \end{tabular}};
		\node[latent,minimum size=\minsize, below=of age, yshift=.3cm] (frontline) {\begin{tabular}{c}
		        Close contact,\\
		        LFD+, ...
		    \end{tabular}};
		\node[latent,minimum size=\minsize, above=of sym] (target) {\begin{tabular}{c}
		        Targeted  \\
		        acoustic\\
		        signature
		    \end{tabular}};
		\node[obs,minimum size=\minsize, left=of covid] (sym) {\begin{tabular}{c}
		        Symptoms  \\
		        (metadata)
		    \end{tabular}};
		\node[obs,minimum size=\minsize, below=of covid] (ciabenroltt) {\begin{tabular}{c}
		        \gls{tt}\\
		        enrolment
		    \end{tabular}};
		\node[obs,minimum size=\minsize, right=of react] (ciabenrolreact) {\begin{tabular}{c}
		        \gls{react}\\
		        enrolment
		    \end{tabular}};
		\node[obs,minimum size=\minsize, right=of ciabenroltt] (getacoustics) {\begin{tabular}{c}
		        Measure  \\
		        acoustics
		    \end{tabular}};
		\node[const, above=of getacoustics, yshift=6.6cm] (via) {};
		\node[const, above=of acoustics, yshift=0cm] (via2) {};

  		\edge[line width=\normalline] {sex} {acoustics}; 
  		\edge[line width=\normalline] {age} {acoustics}; 
  		\edge[line width=\normalline] {sym} {acoustics}; 
  		\edge[line width=\normalline] {age} {sym}; 
  		\edge[line width=\normalline] {covid} {sym}; 
  		\edge[line width=\normalline] {covid} {target}; 
  		\edge[line width=\normalline] {covid} {ciabenroltt}; 
  		\edge[line width=\normalline] {tested} {ciabenroltt}; 
  		\edge[line width=\normalline] {target} {acoustics}; 
  		\edge[line width=\normalline] {sym} {tested}; 
  		\edge[line width=\normalline] {age} {frontline}; 
  		\edge[line width=\normalline] {sex} {frontline}; 
  		\edge[line width=\normalline] {age} {tested}; 
  		\edge[line width=\normalline] {frontline} {tested}; 
  		\edge[line width=\normalline] {react} {ciabenrolreact}; 
  		\edge[line width=\normalline] {ciabenrolreact} {getacoustics}; 
  		\edge[line width=\normalline] {ciabenroltt} {getacoustics}; 
  		\edge[-,line width=\normalline] {getacoustics} {via}; 
  		\edge[-,line width=\normalline] {via} {via2}; 
  		\edge[line width=\normalline] {via2} {acoustics}; 


	\end{tikzpicture}
\end{center}
\vspace{-.25cm}
    
	\caption{Bayesian knowledge graph describing the main features of the recruitment process. The nodes in the graph represent the states of an individual in the population; shaded nodes are observed and non-shaded are latent.}
\label{fig:knowledge_graph_v1}
\end{figure}

\begin{figure}[h!]
    \centering
    \includegraphics[width=\textwidth]{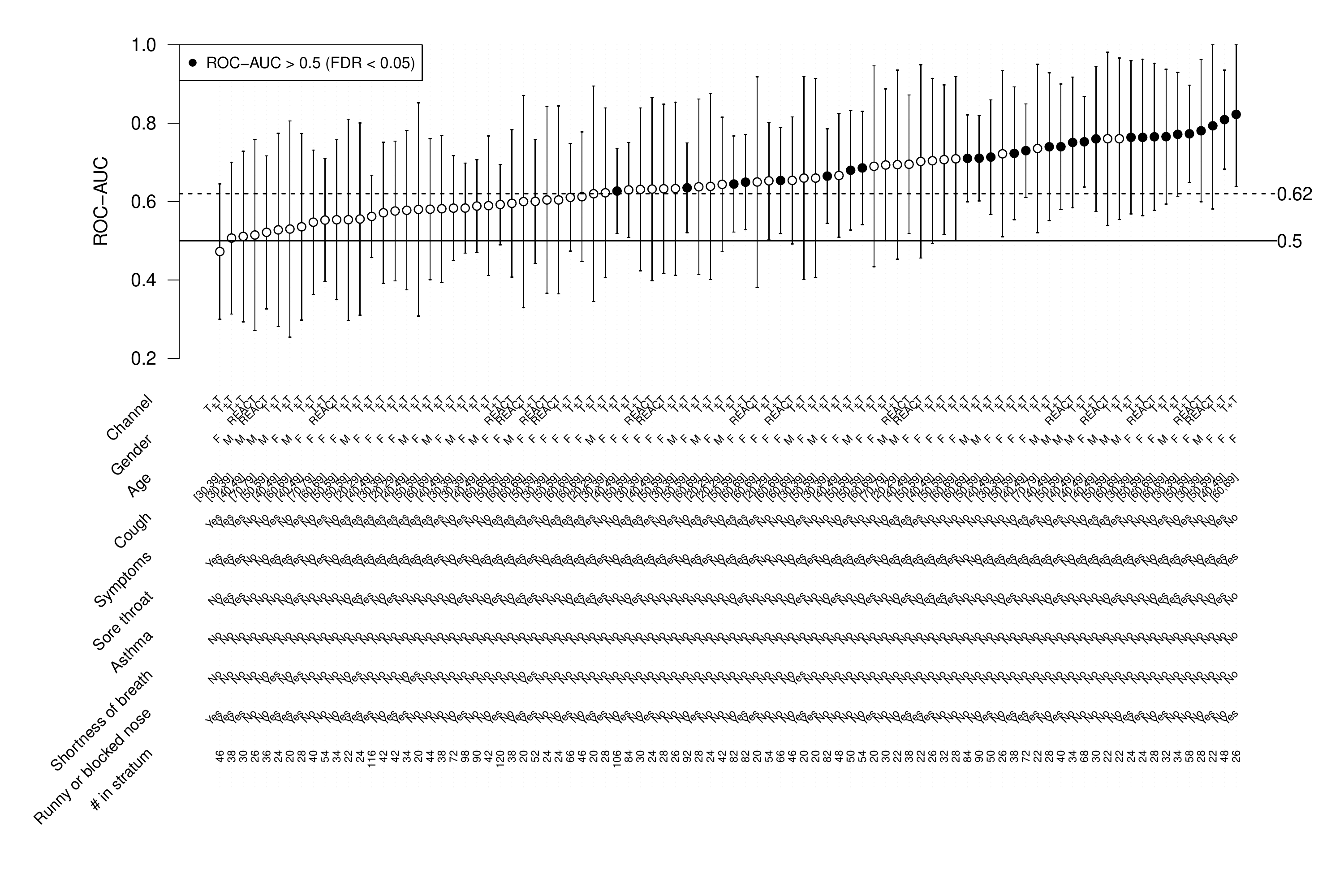}
    \caption{Predictive accuracy within Matched strata. Estimated \gls{roc} in each of 88 strata in the combined Matched and Longitudinal Matched test sets (these are the 88 largest strata in this combined test set, having at least 10 \gls{covidpos} and 10 \gls{covidneg} participants). Upon controlling \gls{fdr} at 5\%, we observe significant differences in predictive scores between \gls{covidneg} and \gls{covidpos} individuals in 28 strata (two-tailed Mann–Whitney U test; significance denoted by filled points), suggesting that the classifier has low but consistent predictive power across a large number of strata. Error bars denote DeLong 95\% confidence intervals; of these CIs, 84 out of 88 (95.4\%) are overlapping with 0.62, consistent with a common value of  \gls{roc}=0.62 across all strata.  
    Details of each stratum are shown in the table below the plot, with the  number of participants in each stratum in the final row of the table. The reference value of \gls{roc}=0.62, representing the estimated global (non-stratified) predictive ability of the SSAST classifier (see Table~\ref{tab:main-results}) is marked with a horizontal dashed line. The value \gls{roc}=0.5, representative of no predictive ability is marked by a solid horizontal line.}
    \label{fig:stratified_roc-auc}
\end{figure}


\begin{figure}[h!]
    \centering
    \includegraphics[width=0.8\textwidth]{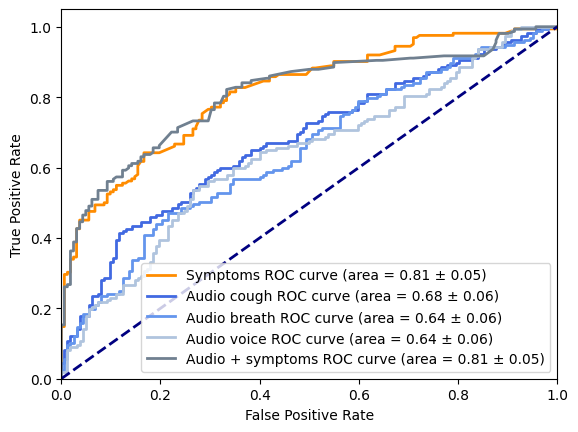}    \caption{\gls{roc} \gls{ssast} performance when trained and evaluated on the COVID-19 sounds publicly available dataset \cite{han_sounds_2022}. Here, our \gls{ssast} model's \gls{roc} exceed those of the \gls{cnn} model of Han \textit{et al.}\cite{han_sounds_2022}, but the difference in \gls{roc} between the methods is small and is compatible with random estimation error, as seen from the wide confidence intervals (attributable to the small test set of size 200): cough (\gls{roc} for \gls{ssast} 0.68 [0.62-0.74] vs \gls{cnn} 0.66 [0.60-0.71]), breath (0.64 [0.58-0.70] vs 0.62 [0.56-0.68]), voice (0.64 [0.58-0.70] vs 0.62 [0.56-0.68]). A simple symptoms checker (\gls{rf}) and a hybrid symptoms-audio are also evaluated for comparison, outperforming both our \gls{ssast} audio-only fit and Han \textit{et al.}'s audio-only \gls{cnn}.}
    \label{fig:cambridge_raw_test_set_results}
\end{figure}

\begin{figure}[h!]
    \centering
    \includegraphics[width=\textwidth]{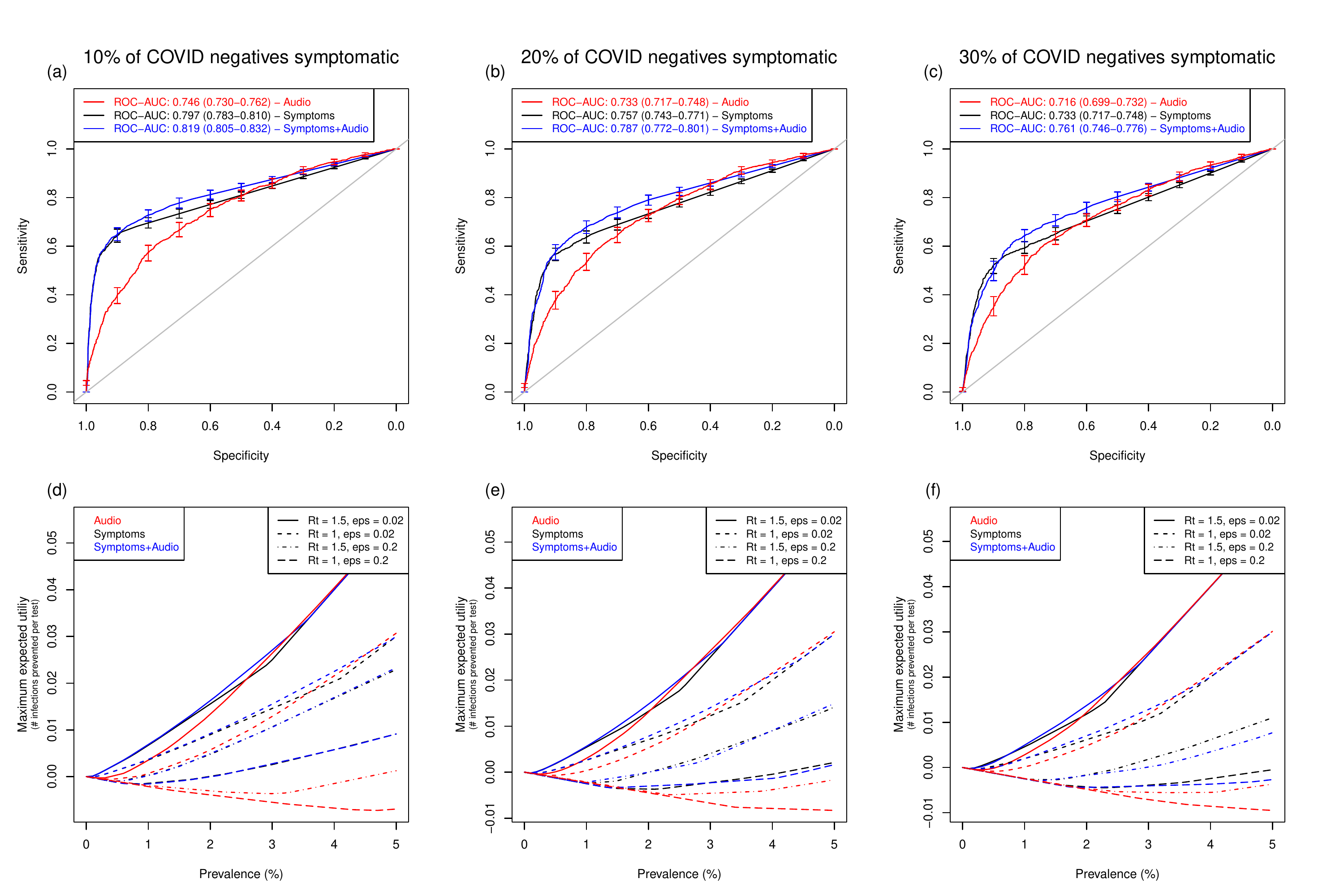}
    \caption{Comparison of sensitivity, specificity, and utility across audio-based and symptoms-based classifiers, as applied in a simulated general populations. The percentage of \gls{covidneg} individuals who are symptomatic in the general population varies between 10\% and 30\% across the three columns of panels (labelled top). (a)-(c) Comparison of ROC curves between the Audio, Symptoms and Symptoms+Audio classifiers;  panel legends show curve colour for each classifier, along with \gls{roc} estimates and 95\% DeLong CIs.
    (d)-(f) Comparison of maximum expected utility across classifiers. Four different utility functions are included, as detailed in the top-right legend (utility function parameters $R_t$, $\varepsilon$ and $\delta$ are defined in Results; in this Figure, $\delta=0.25$). Curves are coloured to indicate Audio, Symptoms or Symptoms+Audio classifiers, as detailed in the top left legend.}
    \label{fig:audio_vs_symptoms_genpop_delta_0.25}
\end{figure}

\begin{figure}[h!]
    \centering
    \includegraphics[width=\textwidth]{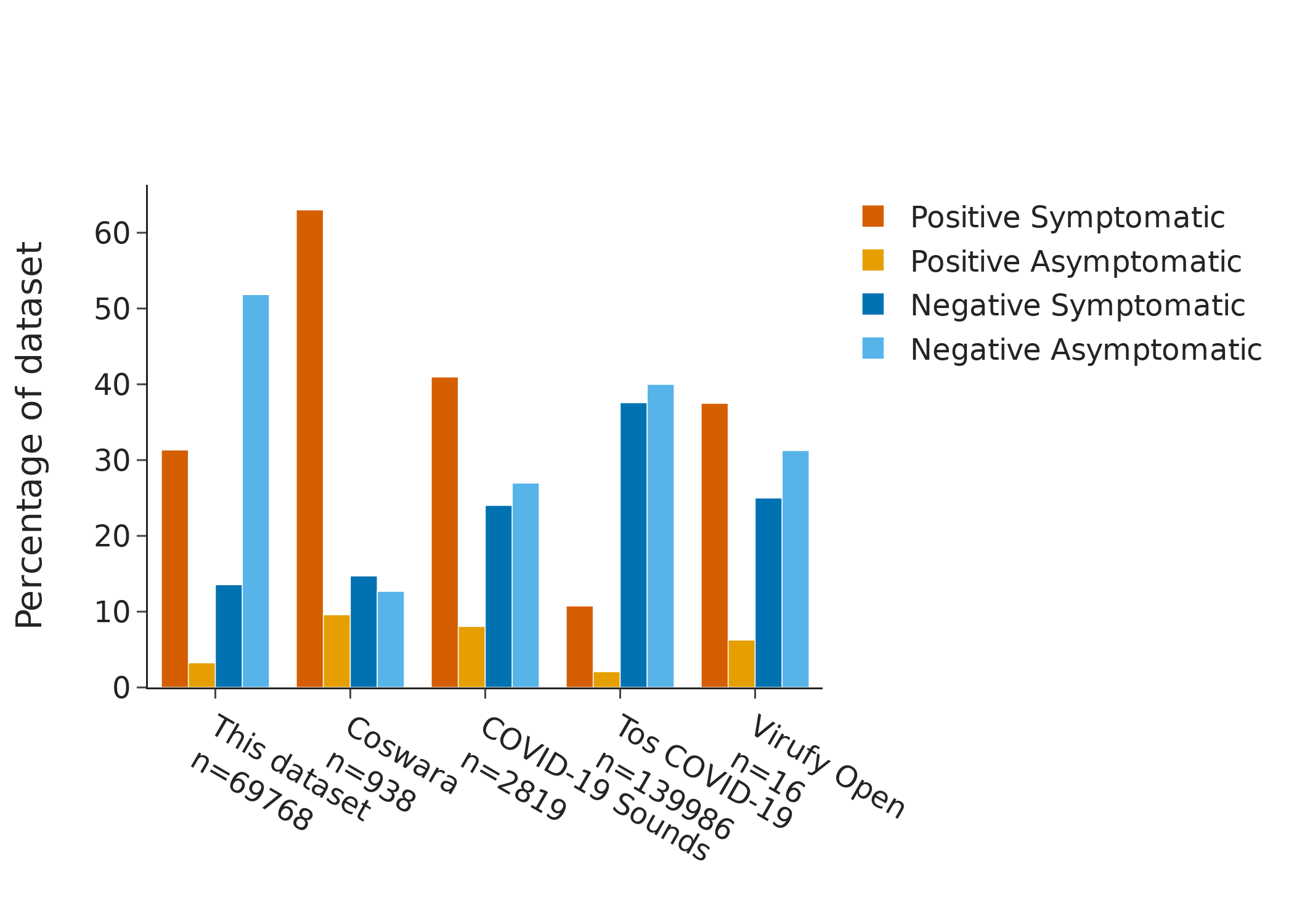}    \caption{Symptomatic vs Asymptomatic for other \gls{covid} datasets when sufficient information is provided. Coswara \cite{sharma2020coswara}, COVID-19 Sounds \cite{han_sounds_2022}, Tos COVID-19 \cite{pizzo_iatos_2021} and Virufy \cite{chaudhari2020virufy}. }
    \label{fig:dataset_symp}
\end{figure}

\begin{figure}[h!]
    \centering
    \includegraphics[width=0.95\textwidth]{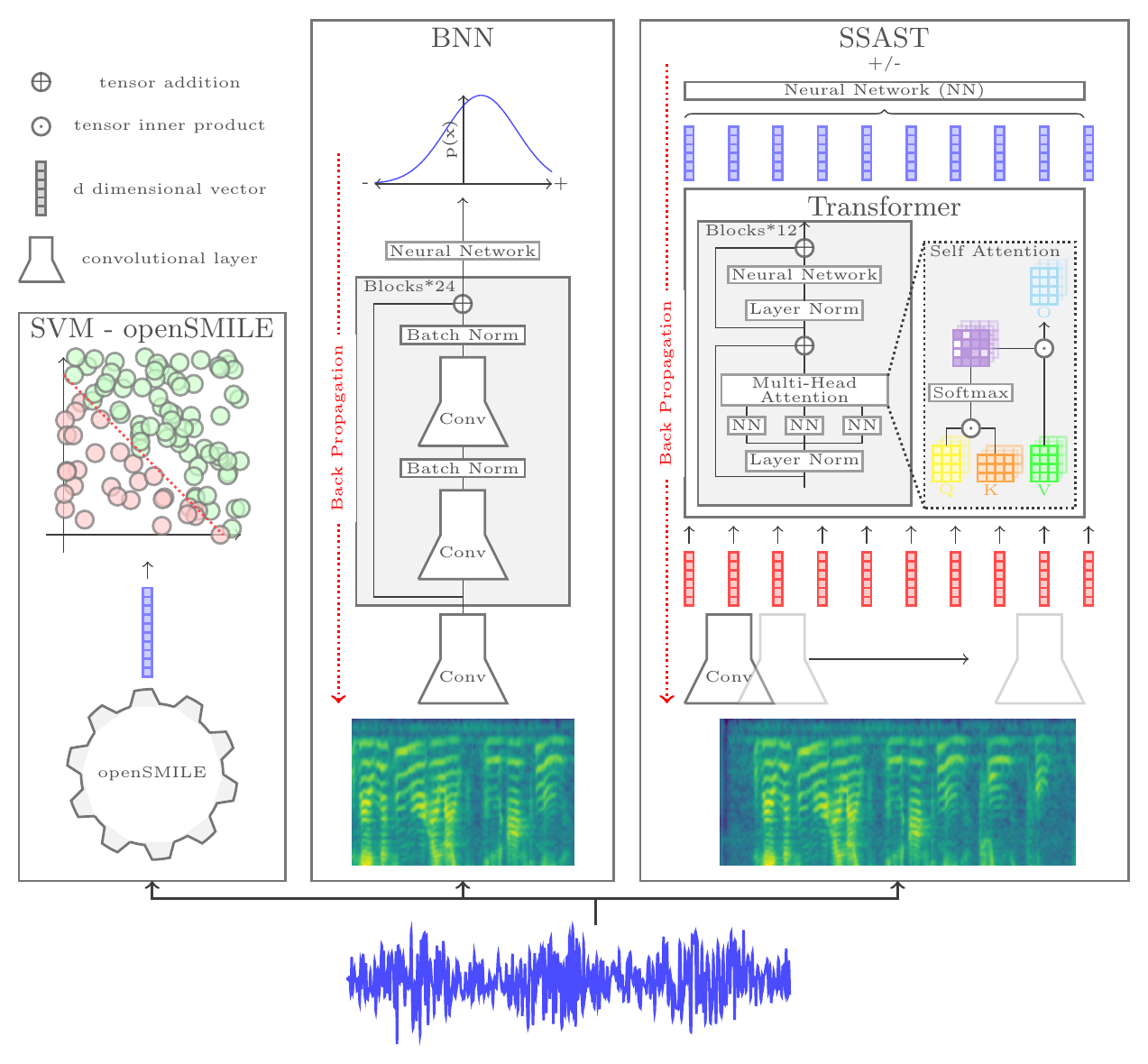}    \caption{Schematic detailing the three separate pipelines implemented to evaluate \gls{abcs}, openSMILE-SVM baseline, the Bayesian Neural Network (BNN) and the Self-Supervised Audio Spectrogram Transformer (SSAST). Both SSAST and BNN first convert the raw audio signal to mel spectrogram space whereas the openSMILE-SVM approach extracts a series of handcrafted audio features.}
    \label{fig:models}
\end{figure}

\begin{figure}[ht]
     \centering
     \includegraphics[width=0.95\textwidth]{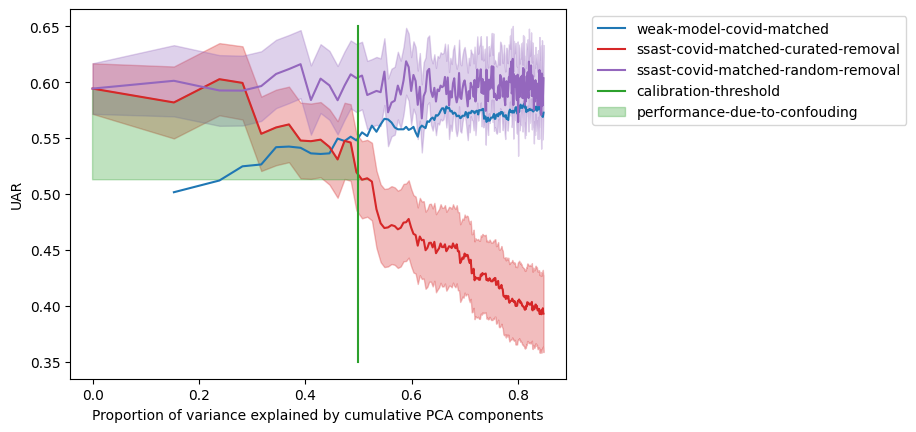}    \caption{Results of the Weak-Robust approach. The blue line represents the \gls{svm} model trained and evaluated on an increasing number of \gls{pca} components of openSMILE vector representations of the audio signal for the Matched \gls{covid} detection from audio task (`weak-model-covid-matched'). Individuals correctly classified by the weak model in the Matched test set are hypothesised to harbour confounding signal, and are removed to create the \textit{curated} Matched test set.
    The red line shows \gls{ssast} performance on this curated Matched test set (`ssast-covid-matched-curated-removal').
    For comparison, we also randomly remove Matched test cases and these results are shown by the purple line (`ssast-covid-matched-curated-removal'). The vertical green line corresponds to the calibration threshold, i.e., the number of PCs for which the weak model achieves \gls{uar} of greater than 80\% on the calibration task.
     The green shaded area corresponds to the drop in \gls{ssast} performance that we attribute to the removal of confounding in Matched test set cases. We note that the drop in performance below random classification is hypothesised to be due to only the ``tricky'' cases remaining (e.g., symptomatic \gls{covidneg}). The 95\% confidence intervals are calculated via the normal approximation method.}
    \label{fig:weak-robust}
\end{figure}

\begin{figure}
\begin{center}
\begin{tabular}{ll}
\subfloat[age]{\includegraphics[width = 1.6in]{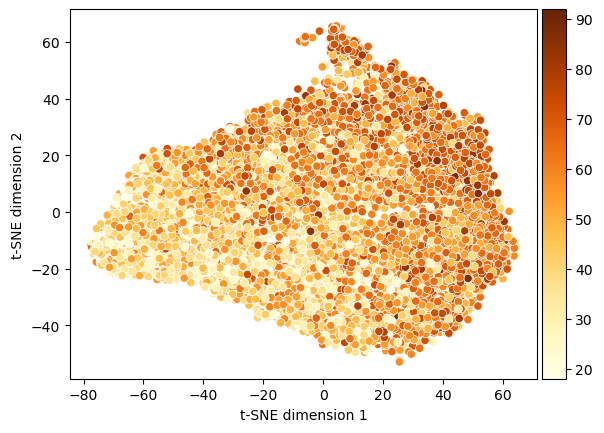}} & \subfloat[matched age]{\includegraphics[width = 1.6in]{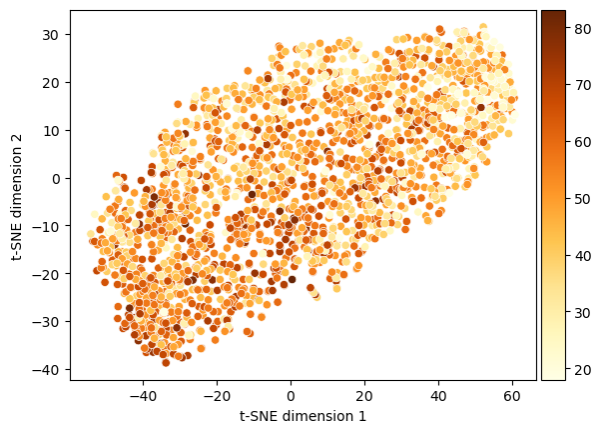}} \\
\subfloat[cough]{\includegraphics[width = 1.6in]{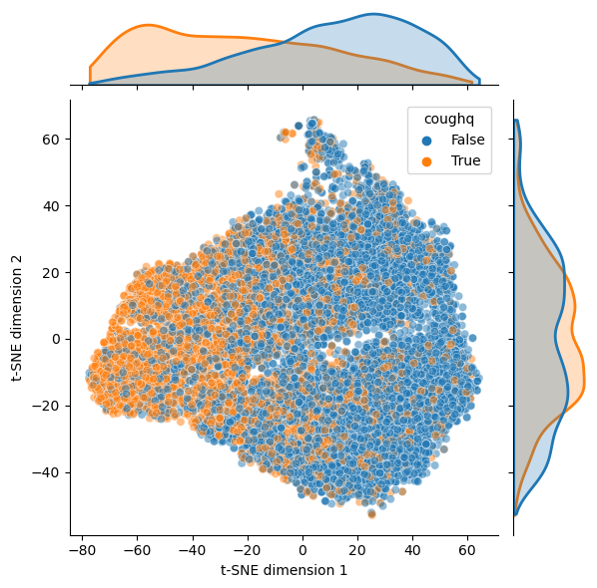}} & \subfloat[matched cough]{\includegraphics[width = 1.6in]{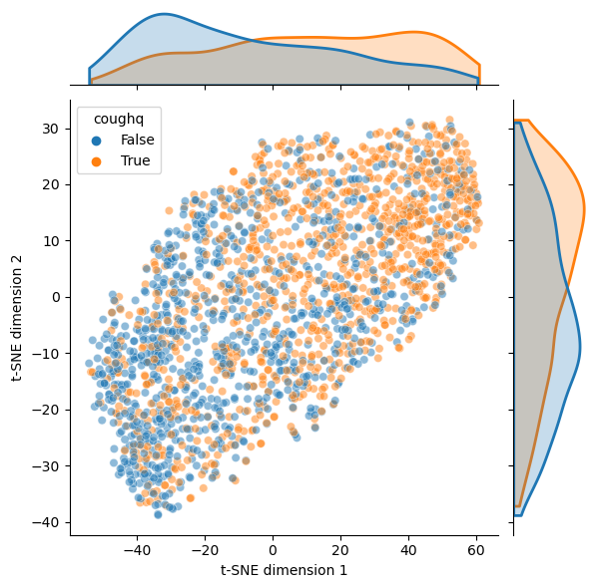}} \\
\subfloat[recruitment source]{\includegraphics[width = 1.6in]{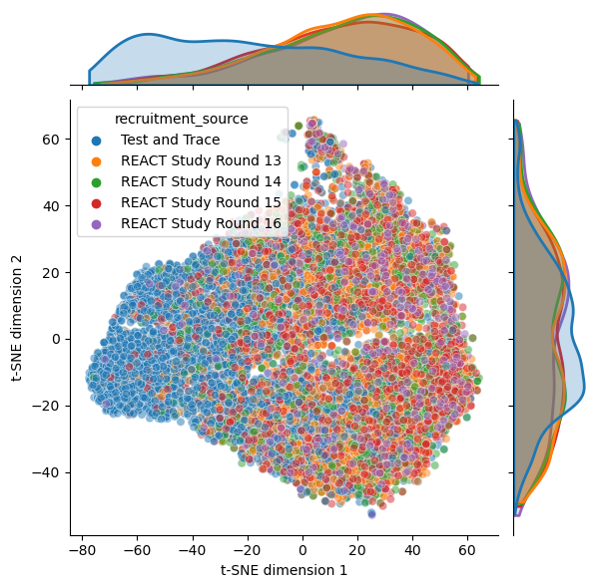}} & \subfloat[matched recruitment source]{\includegraphics[width = 1.6in]{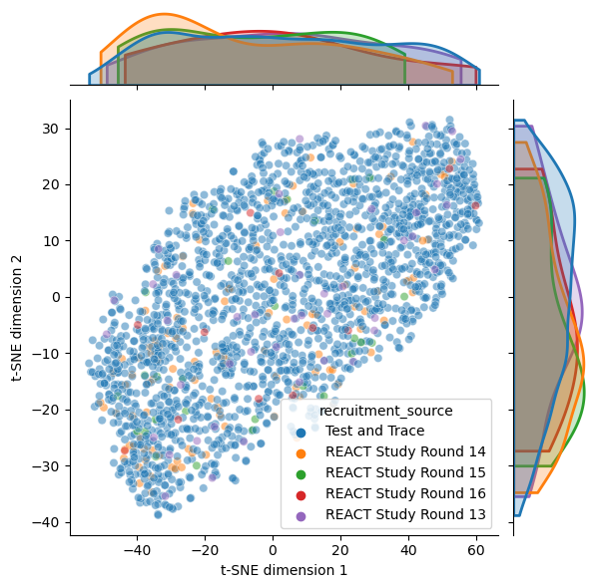}} \\ \subfloat[test result]{\includegraphics[width = 1.6in]{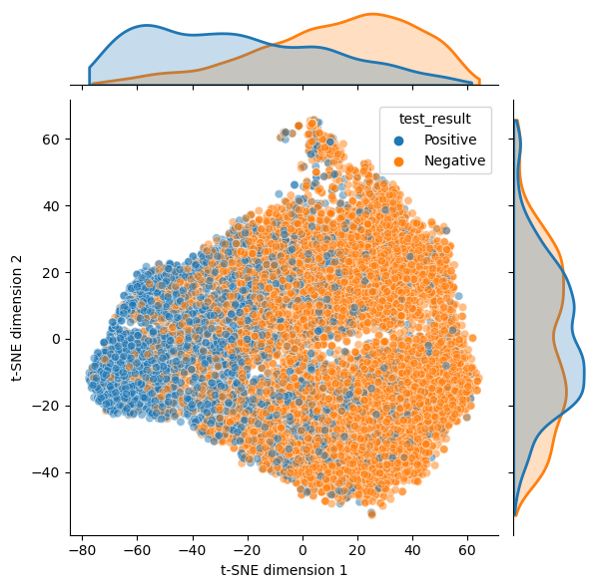}} & \subfloat[matched test result]{\includegraphics[width = 1.6in]{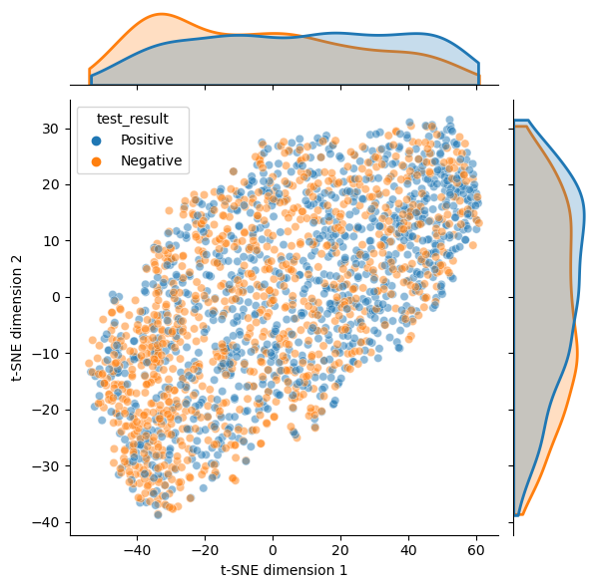}} \\

\end{tabular}
\caption{TSNE plots of the the final layer representations of the SSAST coloured according to confounder and \gls{covid} test result. (a,c,g,e) are Standard test set instances and (b,d,f,h) are Matched test instances. (a-h) are the learnt representations from the same trained model and are a 2-dimensional view of a 768 dimensional space.}
\label{fig:tsne-rep}
\end{center}
\end{figure}

\begin{figure}
     \centering
     \begin{subfigure}[b]{0.495\textwidth}
         \centering
         \includegraphics[trim = {0 142 0 142}, clip,width=\textwidth]{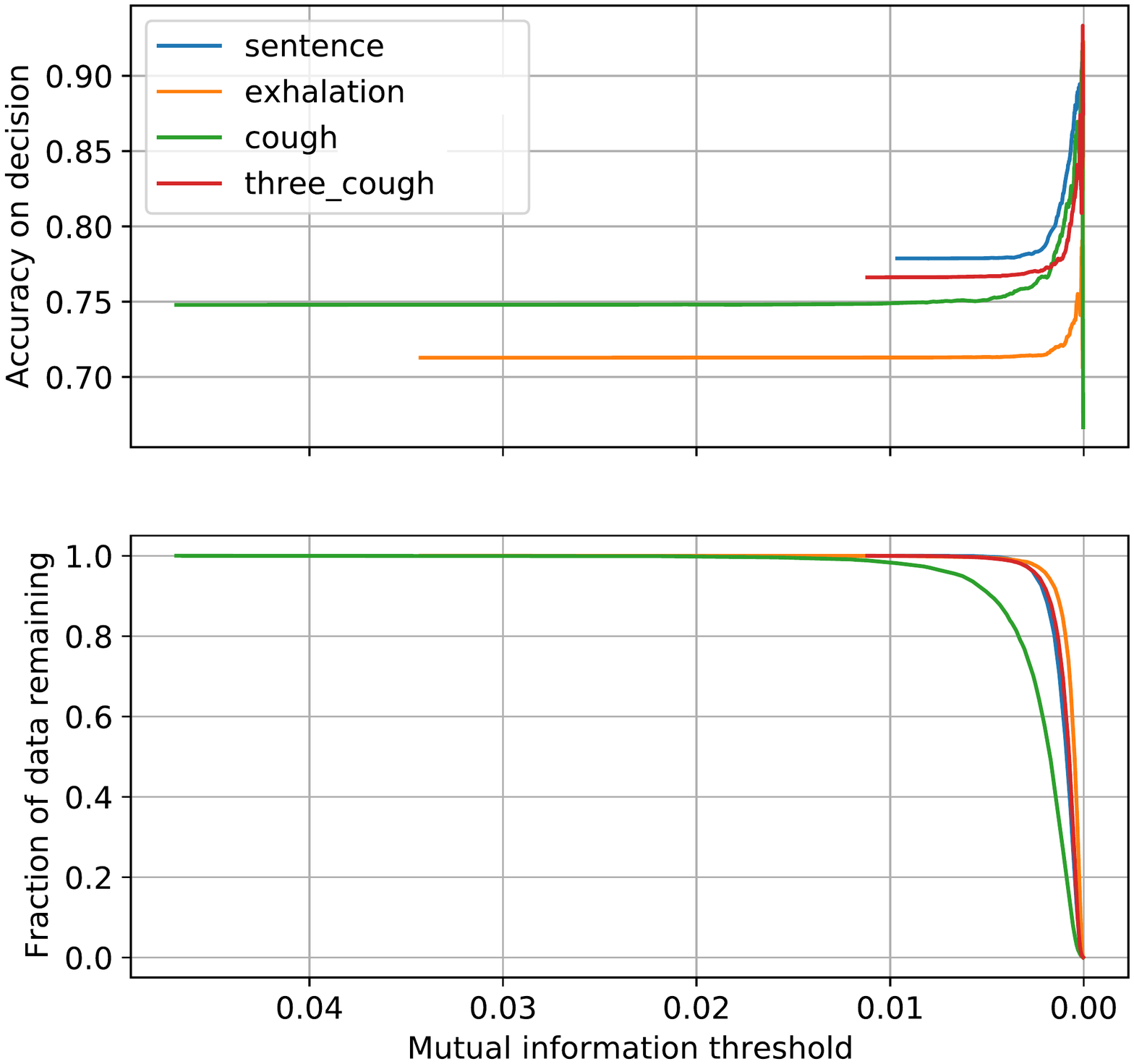}
      \caption{Randomised train-test split MI}

         \label{fig:NaiveMI}
     \end{subfigure}
     \hfill
     \begin{subfigure}[b]{0.495\textwidth}
         \centering
         \includegraphics[trim = {0 142 0 142}, clip,width=\textwidth]{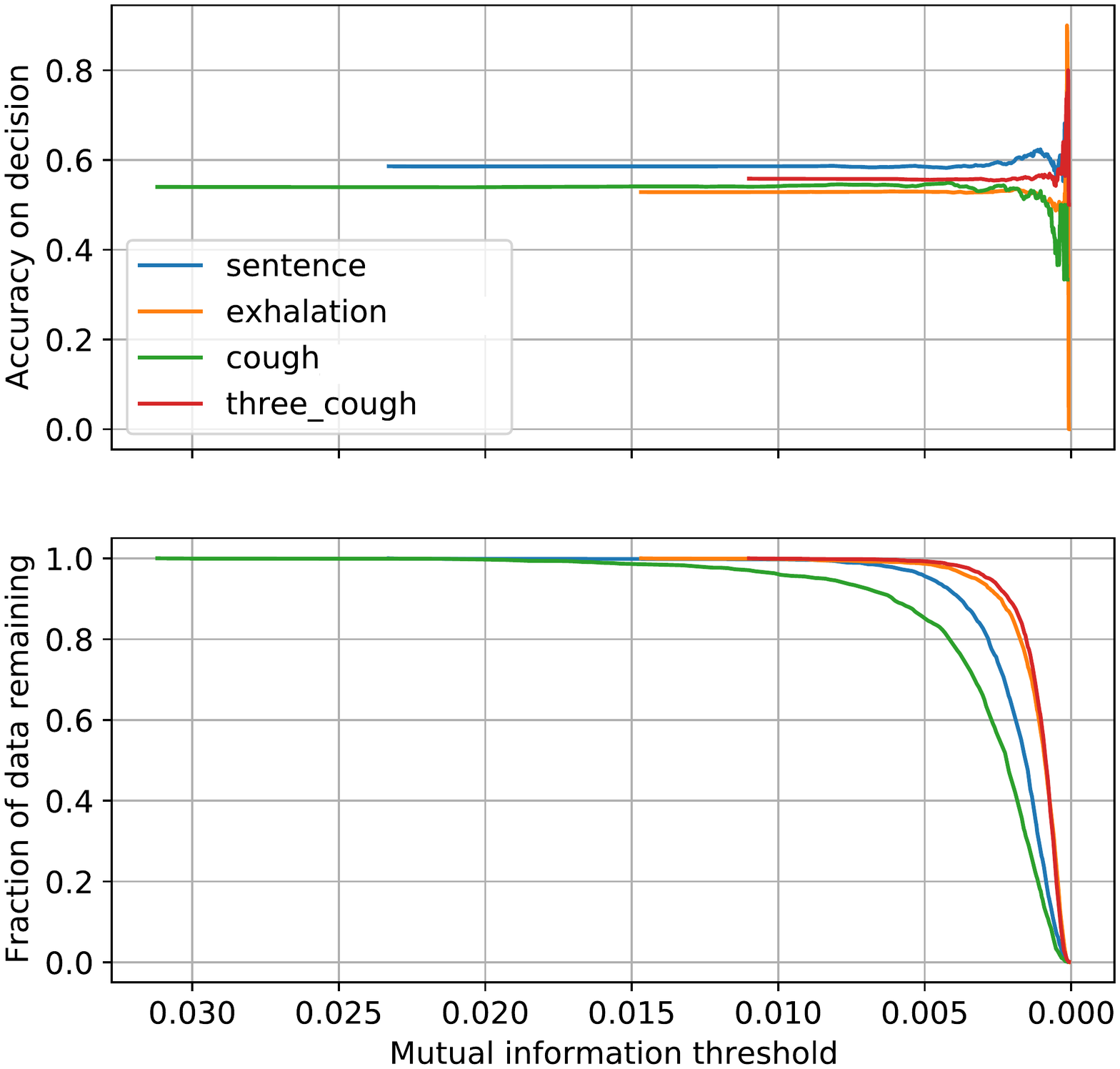}
         \caption{Matched train-test split MI}

         \label{fig:MatchedMI}
     \end{subfigure}

     \begin{subfigure}[b]{0.495\textwidth}
         \centering
         \includegraphics[trim = {0 138 0 157.5}, clip,width=\textwidth]{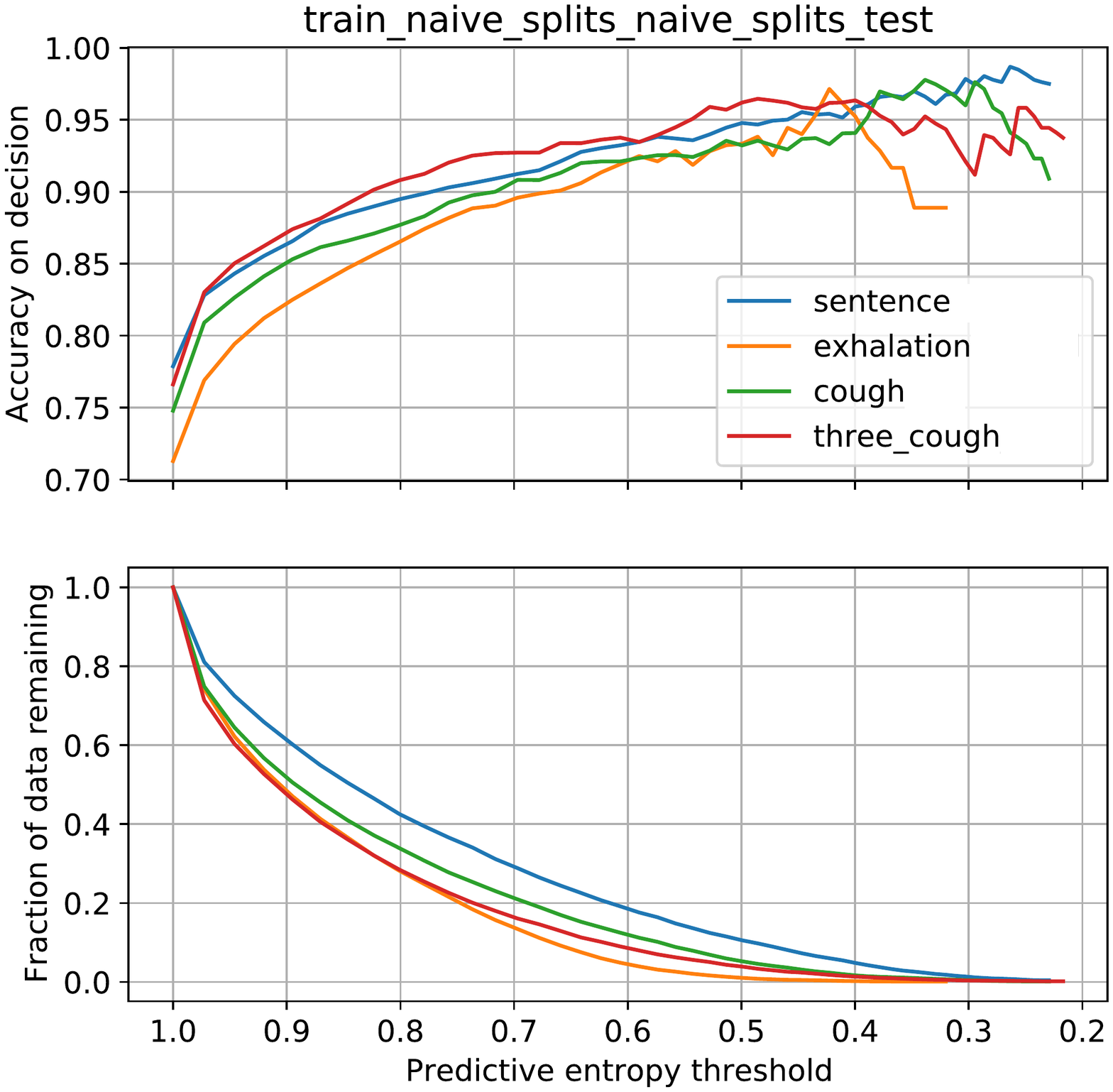}
         \caption{Randomised train-test split PE}
         \label{fig:NaivePE}
     \end{subfigure}
     \hfill
     \begin{subfigure}[b]{0.495\textwidth}
         \centering
         \includegraphics[trim = {0 138 0 158}, clip,width=\textwidth]{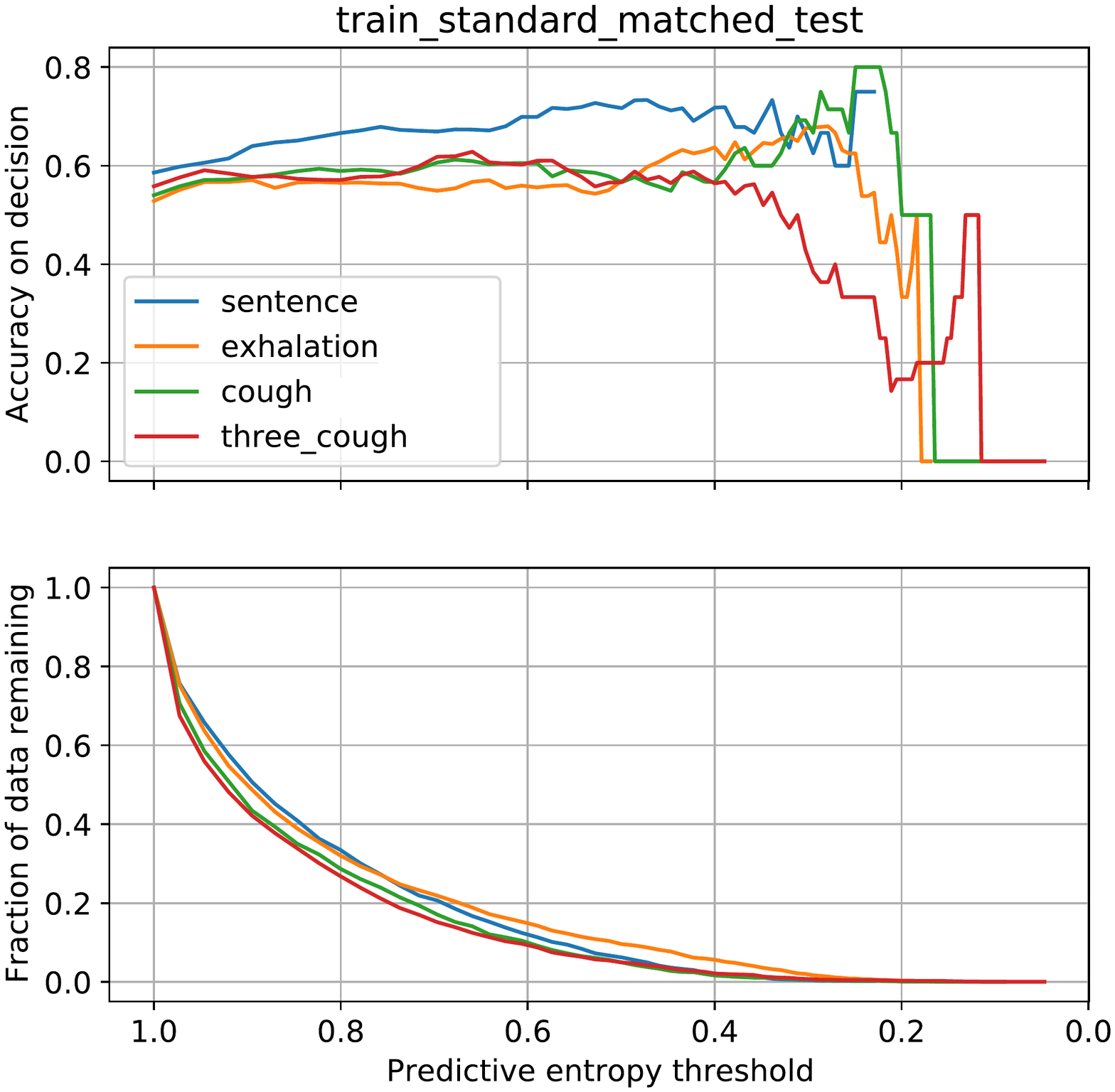}
         \caption{Matched train-test split PE}
         \label{fig:MatchedPE}
     \end{subfigure}
        \caption{The mutual information (MI) and predictive entropy (PE) over the randomised, and matched train-test splits over all audio modalities. The accuracy of the classification is plotted against uncertainty metrics, with the corresponding fraction of data over which the accuracy is calculated given in rows two and four of the figures. The key result is that the BNN is unable to capture a good representation of uncertainty in the case of matched train-test data, however, is able to do so in the randomised train-test split.}
        \label{fig:BNNUncertainty}
\end{figure}

\begin{figure}[h!]
    \centering
    \includegraphics[width=\textwidth]{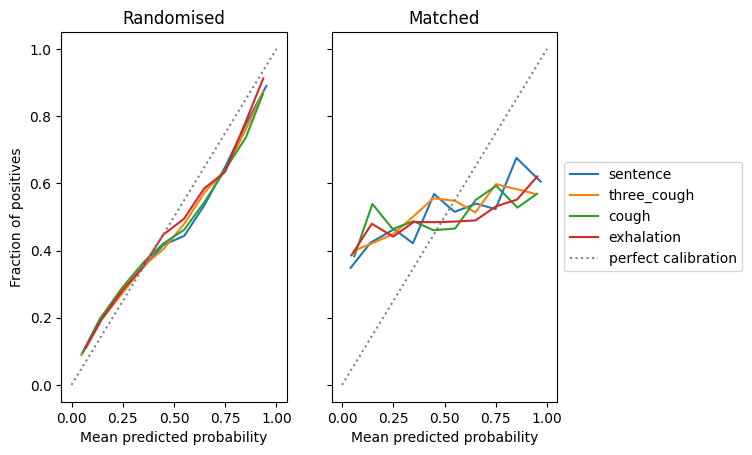}    \caption{Calibration plot for the \gls{ssast} for each respiratory modality for the Randomised and Matched test sets. Here, test case instances are binned according to model predictive probabilities. For each bin, the fraction of \gls{covidpos} is then calculated. For perfect calibration, points will lie along the diagonal\cite{niculescu-mizil2005}.}
    \label{fig:calibration}
\end{figure}

\begin{landscape}
\subsection*{Other Publicly available/highly cited COVID-19 datasets}
\label{sec:otherdatasets}

\begin{table}[!ht]
    \centering
    \begin{tabularx}{\linewidth}{X|X|X|X|X|X|c|c|c}
\specialrule{.2em}{.1em}{.1em}
        Dataset & Positive/Total Participants & Audio Modality & Crowd sourced & COVID-19 Label & Publicly available & \multicolumn{3}{c}{Reported Classification metrics} \\
        \cline{7-9}
        & & & & & & F1 & ROC-AUC & Acc \\ \hline
        Tos COVID-19 \cite{pizzo_iatos_2021} & 25664/139986 & Cough & yes & PCR (19.36\%), Lateral Flow & yes & 0.87 & - & -\\ \hline
        COVID-19 Sounds \cite{han_sounds_2022} & 2106$\dagger$/36116 & Cough, Breathing, Speech & yes & Self-reported & yes & - & 0.71 & \\ \hline
        COUGHVID \cite{orlandic_coughvid_2021} & 1010/20072 & Cough & yes & Self-reported, Clinician annotation & yes & - & - & 0.95$\ddag$ \cite{loey2021covid}\\ \hline
        Covid19-Cough \cite{ponomarchuk2022} & 682/1324 & Cough & yes & PCR (28.85\%), Self-reported & yes & - & - & 0.96 \\ \hline
        Coswara \cite{sharma2020coswara} & 389/2233 & Breathing, Cough, Vowel phonation, Speech & yes & Self-Reported & yes & - & 0.85$\ddag$ \cite{sharma2022} & - \\ \hline
        Virufy \cite{chaudhari2020virufy} & 14/91 & Cough & yes & PCR (75.82\%), Self-reported & yes & 0.93 & - & - \\ \hline
        NoCoCoDa \cite{cohen2020novel} & 10/10 & Cough & yes & Self-reported & yes & 0.77$\ddag$\cite{julio2021} & - & - \\ \hline
        CoughDetect \cite{andreu-perez_generic_2021} & 2339$\dagger$/8380 & Cough & no & PCR (100\%) & no & - & 0.99 & - \\ \hline
        opensigma \cite{laguarta_covid-19_2020} & 2660/\~30000 & cough & yes & no test required & no & - & - & 0.99 \\ 
\specialrule{.2em}{.1em}{.1em}

    \end{tabularx}

    \footnotesize{$\dagger$ recorded samples NOT number of participants}\\
    \footnotesize{$\ddag$ when the dataset paper does not provide classification metrics then scores are taken from other published work which give classification scores of corresponding dataset}\\
    \caption{Non-exhaustive record of COVID-19 respiratory audio datasets. Publicly available datasets and two highly-cited private datasets are shown.}
    \label{tab:other-datasets}
    
\end{table}
\end{landscape}

\clearpage

\section*{Supplementary Note 1}

\subsection*{Exploratory approaches to identify the influence of unmeasured confounders}
\label{sec:algochar}

Matching can be used to control for bias attributable to confounders that are measured. Unmeasured confounders associate with audio signal and \gls{covid} status but are not in the metadata (e.g., unreported symptoms or confounding background noise). The issue with unmeasured confounders is that they can be identified by audio models and lead to artificially inflated and non-generalisable classification performance. We introduce two exploratory approaches aimed at quantifying the effects of unmeasured confounding.

\paragraph{Method 1: Weak-Robust approach.}
We take inspiration from work done in Natural Language Understanding, where weak models are used to propose samples in which bias is likely to be present \cite{utama-etal-2020-towards}. Consider a dataset of pairs of recordings $x \in X$ and \gls{covid} status/labels $y \in Y$. For pairs $(x^i, y^i)$, let the $b(x^i)$ be biased features which might exist in $x^i$ and exhibit association with  $y^i$. We train a low capacity model, $f_b$, designed to be insufficiently expressive to fit the targeted signal. Then, for correctly classified test cases, i.e., for $f_b(x^i) = y^i$, we hypothesise that the instance $x^i$ may contain confounding signal, $b(x^i)$, 
because the model was able to predict the instance correctly despite not being sufficiently expressive to fit the targeted signal. Our approach is to remove individuals for whom the weak model correctly predicts \gls{covid} status, the intention being to remove test set cases which may contain confounding signal. 


We create weak models through training SVM models on the first $k$ Principal Components (PCs) of the openSMILE feature representions. The smaller the value of $k$, the weaker the model. Our approach requires us to design the weak model so as it lacks the capacity to fit the true targeted signal. To do this, we introduce two novel steps. First a calibration step, where we evaluate the weak model on a similar and known easier task than \gls{covid} detection, namely classifying between coughing and laughing sounds. This task was constructed from the established ESC-50 dataset \cite{piczak2015dataset}. This calibration task was chosen as it was also a classification task in human respiratory space. We hypothesise that if the weak model cannot perform well at this calibration task it is too weak to identify \gls{covid} audio features. Second, we project the \gls{covidpos} signal to the null space through projecting both the \gls{covidpos} and \gls{covidneg} cases onto the first $k$ PCs of only the \gls{covidneg} cases. Good classification performance in this space is likely to be due to variation which exists in the \gls{covidneg} population and not the \gls{covidpos} population.

As detailed in Figure~\ref{fig:weak-robust}, as we remove the cases which we hypothesise to contain confounding signal, the \gls{ssast} performance on the Matched test set quickly drops to a random prediction (UAR=0.51 [0.47 -- 0.55]). We note that further removal of Matched test instances beyond the calibration threshold point leads to a drop in performance below 0.5. This is to be expected as we are removing the cases which the weak model is able to classify well, leaving behind only the miss-classified cases, cases the \gls{ssast} evidently also struggles to correctly classify. As $k$ has exceeded the point at which the weak model performs well at the calibration task, this further drop in performance could either be due to the weak model identifying instances which truly contain the signal for \gls{covid} or more bias.

\paragraph{Algorithm Weak-Robust}
\begin{enumerate}
    \item Begin with all individuals in Curated Matched test set
    \item For \# PCs $k = 1:K$
    \begin{enumerate}
        \item Project \gls{covidpos} and \gls{covidneg} openSMILE features onto the first $k$ principal components of the \gls{covidneg} cases
        \item Train and evaluate prediction metrics in PCA-reduced Matched test set
        \item Remove correctly classified individuals from Curated Matched test set
        \item Evaluate \gls{ssast} model on Curated Matched test set
    \end{enumerate}
    \item The cumulative drop in performance can be attributed to removing test instances which contain identifiable bias.
    \item How do we know that we are not removing cases where the \gls{covid} signal is particularly easy to identify? Compare weak model performance to a calibration task which we know is easier that \gls{covid} detection from audio. Weak models which lack the capacity to perform well at this calibration task are hypothesised to be too weak to learn the true signal and therefore must be leveraging bias.
\end{enumerate}

\paragraph{Method 2: Nearest neighbour approach.}

This next method explores how much of the learnt signal for \gls{covid} exists in a \gls{covidneg} population. If the learnt \gls{covid} signal is to have the desired properties of P1 and P3 (defined in Results - Defining the acoustic target for audio-based \gls{covid} screening) this signal should not be expressive in a space defined purely by \gls{covidneg} recordings.

\paragraph{Algorithm NN}
\begin{enumerate}
    \item For each \gls{covidpos} participant $i\in\{i:y_i=1\}$ in Matched test set, identify the audio-nearest-neighbour \gls{covidneg} individual $i^\prime$ in the Matched test set $$i^\prime = \underset{i^\prime}{\text{argmin}}\  d(x_i, x_{i^\prime})\ .$$ Audio nearest-neighbour distance $d(x_i, x_{i^\prime})$ is defined, e.g., by Euclidean or Manhattan distance in K-dimensional score space of PCA on openSMILE features.
    \item In the Matched test set, replace each \gls{covidpos} participant's audio with the NN \gls{covidneg} participant's audio:
    $$x_i \leftarrow x_{i^\prime}\ \forall\ i\in\{i:y_i=1\}$$
    \item Evaluate predictive accuracy metrics in Matched test set; if we still obtain \gls{roc}=0.60 \emph{and} the signal is common in \gls{covidneg} (i.e., many distinct \gls{covidneg} are chosen as NN) we attribute the signal to unmatched/unmeasured confounders.
\end{enumerate}

\clearpage
\section*{Supplementary Note 2}

\subsection*{Trends in learnt representations.}
\label{sec:trends}
Analysing the model representation distributions can reveal aspects of the features learnt by the model. We take the trained \gls{ssast} model and extract the final layer representations of the Standard and Matched test sets. The final layer representations are shown schematically in Figure~\ref{fig:models} by the mean pooling of the blue vectors in the \gls{ssast} model. These 768 dimensional vectors are then compressed to two dimensions using \gls{tsne} to allow for visual inspection. Figure~\ref{fig:tsne-rep} compares the learnt representations between the Standard test set and the Matched test set. If we first inspect Figure~\ref{fig:tsne-rep} (g) the model is clustering by \gls{covid} test result. This initially suggests that that the model has identified a signal for \gls{covid}. However, if we look at Figure~\ref{fig:tsne-rep} (c) this same clustering pattern correlates with whether or not the participant is displaying a cough as a symptom or not. This is also the case, to an extent, for recruitment source (Figure~\ref{fig:tsne-rep} (e)) and age distribution (Figure~\ref{fig:tsne-rep} (a)). Now, if we inspect the learnt representations of the Matched test set for Figure~\ref{fig:tsne-rep} (b), (d) and (f) (age, cough and recruitment source respectively) we still see clustering of these confounders. But crucially, as shown in Figure~\ref{fig:tsne-rep} (h), we no longer see the same clustering with \gls{covid} status. This suggests that the model was not leveraging a \gls{covid} acoustic signal, rather other confounding signals which correlated with \gls{covid} status.

\clearpage
\section*{Supplementary Note 3}
\subsection*{Uncertainty quantification with \glspl{bnn}}
\label{sec:BNNUncertainty}
\subsubsection*{Uncertainty metrics}

Having trained a \gls{bnn}, we have a collection of model weights $ 
\{\boldsymbol{\omega}\}_{s=1}^{S}$ for our MC inference scheme of $S$ dropout samples. We want the output of our model $\mathbf{y^*}$ to display its confidence in a class label, $c$. We wish to further examine to what degree the model's predicted uncertainties are consistent, and correct.
For the MC approach there are multiple outputs, where each output corresponds to a different weight sample, $\boldsymbol{\omega}^{(s)}$. We use the definition of the posterior predictive entropy $\tilde{{H}}$ as in \cite{pmlr-v70-gal17a}:
\begin{equation}
    \tilde{H} = - \sum_c \tilde{p}_c \log \tilde{p}_c, \quad \mathrm{where} \quad \tilde{p}_c = \frac{1}{S}\sum_s p_c^{(s)} .
\end{equation}
The predictive entropy (PE) is thereby a single value measure that can indicate the confidence of a model in its prediction. 
In addition, to take into account the origin of the uncertainty (i.e., is it the model that is unsure, or is the data simply noisy), we can decompose the uncertainty components to distinguish the model uncertainty from the data uncertainty. We achieve this with the mutual information (MI) \cite{houlsby2011bayesian}, $ I(\mathbf{y}^* ,\boldsymbol{\omega})$ between the prediction $\mathbf{y}^*$ and the model posterior over $\boldsymbol{\omega}$:
 \begin{equation}
     I(\mathbf{y}^* ,\boldsymbol{\omega}) = \tilde{{H}} -  \mathbb{E} [{H}] ,
 \end{equation}
where we introduce the expectation over the entropy $\mathbb{E} [{H}]$ with respect to the parameters:
\begin{equation}
    \mathbb{E} [{H}] = \frac{1}{S} \sum_s h(\boldsymbol{\omega}_s), \quad \mathrm{where} \quad h(\boldsymbol{\omega}) = - \sum_c p_c(\boldsymbol{\omega}) \log p_c(\boldsymbol{\omega}) .
\end{equation}
The MI will measure how much one variable, e.g., $\boldsymbol{\omega}$, tells us about the other random variable, $\mathbf{y}^*$. If $I(\mathbf{y}^* ,\boldsymbol{\omega}) = 0$, then that tells us that 
$\boldsymbol{\omega}$ and $\mathbf{y}^*$ are independent, given the data.
In the scenario where the predictions completely disagree with each other for a given $\mathbf{x}^*$, for each $\boldsymbol{\omega}_s$ drawn from the posterior, we get very different predictions. This informs us that $\mathbf{y}^*$ is highly dependent on the posterior draw and thus $ I(\mathbf{y}^* ,\boldsymbol{\omega})$ will tend to its maximum value. However, if $\mathbf{y}^* $ is the same for all $\boldsymbol{\omega}_s \sim p(\boldsymbol{\omega}\vert \mathbf{Y,X}$), then the different draws from the posterior distribution have no effect on the predictive distribution and therefore the mutual information between the two distributions is zero, as $\mathbb{E} [{H}] = h(\boldsymbol{\omega}) = \tilde{{H}}$ (they are independent). \cite{cobb2020practicalities}.
\subsubsection*{Results}

Figure \ref{fig:BNNUncertainty} shows how thresholding according to the PE and MI affects the accuracy on decision, defined as the fraction of correct classifications made, for all audio modalities. For each sub Figure, the lower subplot also shows the corresponding fraction of data over which the classifications are made according to the uncertainty threshold.

The key result is that, for the randomised split, both the PE and MI of Figure \ref{fig:NaiveMI} and \ref{fig:NaivePE} show a steady improvement in accuracy. For example, the sentence modality shows an improvement from 0.77, without thresholding over the full data split, to 0.93 at a PE threshold of 0.6 over the remaining 20\% of the randomised test data. This shows consistent model behaviour and suggests that the uncertainty is well quantified. It has been shown that for BNNs, stronger predictive performance also results in better uncertainty quantification and improved robustness \cite{carbone2020robustness}.

In contrast, when using the matched test dataset, Figures \ref{fig:MatchedMI} and \ref{fig:MatchedPE} show that the model uncertainty does not exhibit this desirable behaviour, suggesting that the modelling process has failed to identify a satisfactory mapping between input and output. This shows further evidence that the models learnt were not able to learn a good representation of their uncertainty, thus highlighting that a causal link between acoustic biomarkers and \gls{sarscov2} was not learnt
adequately.

\end{document}